\documentclass[useAMS,usenatbib]{mn2e}
\usepackage{txfonts,graphicx,natbib}
\bibpunct{(}{)}{;}{a}{}{,}
\begin{document}




\title[Type Ibn (SN 2006jc-like) events]{Massive stars exploding in a He-rich circumstellar medium.  \\
    I. Type Ibn (SN 2006jc-like) events}

\author[Pastorello et al.]{A. Pastorello$^{1}$
\thanks{a.pastorello@qub.ac.uk}, 
S. Mattila$^{1,2}$, L. Zampieri$^{3}$, M. Della Valle$^{4,5}$, S. J. Smartt$^{1}$, S. Valenti$^{6,7}$, \and
I. Agnoletto$^{3,8}$, S. Benetti$^{3}$, C. R. Benn$^{9}$, 
D. Branch$^{10}$,  E. Cappellaro$^{3}$,
M. Dennefeld$^{11}$, \and J. J. Eldridge$^{1,12}$, A. Gal-Yam$^{13}$, A. Harutyunyan$^{3,8}$, I. Hunter$^{1,9}$,
H. Kjeldsen$^{14}$, Y. Lipkin$^{15}$,  \and
P. A. Mazzali$^{16,17}$, P. Milne$^{18}$, H. Navasardyan$^{3}$,
E. O. Ofek$^{19}$, E. Pian$^{16}$, O. Shemmer$^{20}$, \and  S. Spiro$^{1,3}$,
R. A. Stathakis$^{21}$, S. Taubenberger$^{17}$, M. Turatto$^{3}$,
H. Yamaoka$^{22}$\\
$^{1}$ Astrophysics Research Centre, School of Mathematics and Physics, Queen's University Belfast, Belfast BT7 1NN, United Kingdom\\
$^{2}$ Tuorla Observatory. University of Turku, V\"ais\"al\"antie 20, FI-21500, P\"ukki\"o, Finland\\
$^{3}$ INAF Osservatorio Astronomico di Padova, Vicolo dell' Osservatorio 5, I-35122 Padova, Italy\\
$^{4}$ INAF - Osservatorio Astrofisico di Arcetri, Largo E. Fermi 5, I-50125 Firenze, Italy\\
$^{5}$ International Center for Relativistic Astrophysics Network,, Piazzale della Repubblica 10,I-65122 Pescara, Italy\\
$^{6}$ European Southern Observatory (ESO), Karl-Schwarzschild-Str. 2, D-85748 Garching bei M\"unchen, Germany\\
$^{7}$ Dipartimento di Fisica, Università di Ferrara, via Saragat 1, I-44100 Ferrara, Italy\\
$^{8}$ Dipartimento di Astronomia, Università di Padova, Vicolo dell'Osservatorio 2, I-35122 Padova, Italy\\
$^{9}$ Isaac Newton Group of Telescopes, Apartado 321, E-38700 Santa Cruz de La Palma, Spain\\
$^{10}$ Homer L. Dodge Department of Physics and Astronomy, University of Oklahoma, Norman, OK 73019-2061, United States\\
$^{11}$ Institut d'Astrophysique de Paris, CNRS, and Université P. et M. Curie, 98bis Bd Arago, F-75014 Paris, France\\
$^{12}$ Institute of Astronomy, The Observatories, University of Cambridge, Madingley Road, Cambridge CB3 0HA, United Kingdom \\
$^{13}$ Benoziyo Center for Astrophysics, Weizmann Institute of Science, 76100 Rehovot, Israel\\
$^{14}$ Danish Asteroseismology Centre, Institut for Fysik og Astronomi, Aarhus Universitet, Ny Munkegade, Bygn. 1520, 8000 Aarhus
C., Denmark\\
$^{15}$ School of Physics and Astronomy and Wise Observatory, Tel Aviv University, 69978 Tel Aviv, Israel\\
$^{16}$ INAF Osservatorio Astronomico di Trieste, Via Tiepolo 11, I-34131 Trieste, Italy\\
$^{17}$ Max-Planck-Institut f\"ur Astrophysik, Karl-Schwarzschild-Str. 1, D-85741 Garching bei M\"unchen, Germany\\
$^{18}$ Department of Astronomy and Steward Observatory, University of Arizona, Tucson, AZ 85721, United States\\
$^{19}$ Division of Physics, Mathematics, and Astronomy, California Institute of Technology, Pasadena, CA 91125, United States\\
$^{20}$ Department of Astronomy and Astrophysics, 525 Davey Laboratory, Pennsylvania State University, University Park, PA 16802,
USA\\
$^{21}$ Anglo-Australian Observatory, PO Box 296, Epping, NSW 1710, Australia\\
$^{22}$ Department of Physics, Kyushu University, Fukuoka 810-8560, Japan\\
}

\date{Accepted .....; Received .....; in original form .....}

\maketitle

\label{firstpage}


\begin{abstract}
We present new spectroscopic and photometric data of the type Ibn supernovae 2006jc, 
2000er and 2002ao. We discuss
the general properties of this recently proposed supernova family, which also includes SN 1999cq. 
The early-time monitoring of SN 2000er traces the evolution of this class of objects during the first few 
days after the shock breakout. An
overall similarity in the photometric and spectroscopic 
evolution is found among the members of this group, which would be unexpected if the energy
in these core-collapse events was dominated by the interaction between supernova ejecta 
and circumstellar medium. Type Ibn supernovae appear to be rather normal type Ib/c supernova 
explosions which occur within a He-rich circumstellar environment.
SNe Ibn are therefore likely produced by the explosion of Wolf-Rayet progenitors 
still embedded in the He-rich material lost by the star in recent mass-loss episodes, 
which resemble known luminous blue variable eruptions. 
The evolved Wolf-Rayet star could either result from the evolution of a very massive star or be
the more evolved member of a massive binary system.
We also suggest that there are a number of arguments in favour of a type Ibn classification for
the historical SN 1885A (S-Andromedae), previously considered as an
anomalous type Ia event with some resemblance to SN~1991bg. 
 
\end{abstract}


\section{Introduction}

The most massive stars, i.e. those with initial masses larger than 30$M_\odot$, 
are thought to end their lives with the core-collapse explosion of a Wolf-Rayet 
(WR) star. Before this stage, many of them undergo a short
period of instability called the luminous blue
variable (LBV) phase. This phase probably lasts of order 10$^{4}$-10$^{5}$ years
\citep{mae87,hum94,boh97,vink02,smi08}, during which they lose mass through 
recurrent mass-loss episodes. Most LBVs undergo fairly minor variability 
(typically below 1 magnitude) which is commonly called the S Doradus phase. 
But occasionally LBVs experience giant outbursts in which their luminosity rapidly 
increases and they can reach absolute visual 
magnitudes around $-14$ \citep{dav97,mau05,vd06}, accompanied by the ejection of  
a large portion of their hydrogen envelope \citep{smi6}. An overview of our current 
knowledge of LBVs can be found in \citet{vge01}.
At the time of explosion, which is expected to occur a few $\times$ 10$^5$ years
after the last major LBV outburst \citep[see e.g.][]{heg03,eld04}, the mass of the WR star is
expected to be of the order of 15-20$M_\odot$. These stars may eventually explode as type Ib supernovae (SNe
Ib) if they have lost the whole H envelope, or as type Ic supernovae (SNe Ic) 
if the stars are also stripped of their He mantles \citep[for a comprehensive description 
of the different SN types see][]{fil97,tura07}.

However, recent work \citep{gal07,kot06,smi06} has suggested the possibility 
that some very massive stars may explode already {\sl during} the LBV phase, producing luminous 
supernovae (SNe). Previously, a post-LBV channel was proposed by \citet{sal02} to explain the observed
properties of the type IIn SN~1997eg. 
In the proposed scenario, LBVs may eventually produce SNe which show high luminosity, 
blue colour, narrow H$\alpha$ line in emission, and very slow spectro-photometric evolution. Their spectra are dominated by very narrow
($\lesssim$1000 km s$^{-1}$), prominent H emission lines which sometimes show  complex, multicomponent 
profiles. These objects would therefore belong to the so-called {\sl type IIn} SN class \citep{sch90}. 
The observed properties of these events are consistent with a scenario in
which a significant fraction of the kinetic energy of the SN ejecta is converted
into radiation via interaction with a dense, H-rich circumstellar medium (CSM).

\begin{figure}
\includegraphics[width=8.5cm]{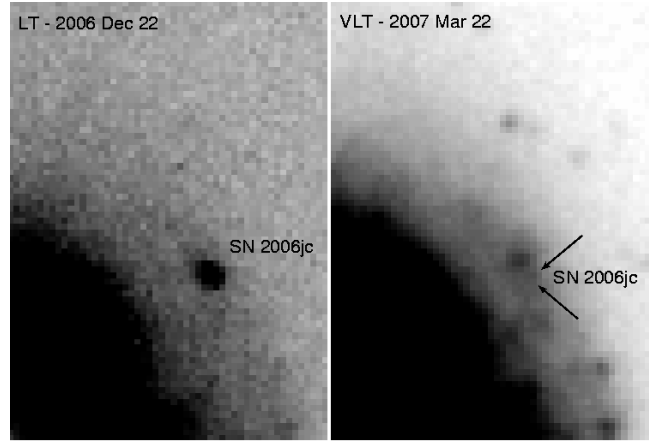}
\caption{Late R-band images of SN~2006jc. Liverpool Telescope image obtained on
December 22, 2006 (left) and a very late VLT image (March 22, 2007; right). In this image, the
SN is only marginally detectable and it is very close to a luminous star-forming region. North is down, East is to the left.
\label{fig1}}
\end{figure}

A new clue for the death of the most massive stars has recently been 
proposed for the explosion of the peculiar SN 2006jc \citep{ita06,pasto07,fol07,smi07,tom07,seppo07}.
Extensive data sets spanning many wavelength regions have been also 
presented by \citet{imm07b,anu07,elisa07,sak07}. 
\cite{pasto07} showed the amazing discovery that a major outburst, 
reaching M$_R = -14.1$ (i.e. an absolute magnitude which is comparable to that 
of LBV eruptions), occurred only two years before the explosion of SN 2006jc, at exactly the same 
position as the SN and inferred that the two events must be physically related. 
The physical connection between the 
two transients is supported by analysis of the SN spectra.  
These are indeed blue and dominated by relatively narrow ($\sim$2200 km s$^{-1}$) 
He I lines in emission \citep[hence the new classification as {\sl type Ibn},][]{pasto07}, 
likely indicating a type Ib/c SN embedded within a dense, massive He-rich envelope.
Unlike SNe IIn, the photometric evolution of this object is extremely rapid, while
the spectra do not change as quickly as the luminosity. 
The best studied object of this type is SN~2006jc, but  three additional events
which appear to be spectroscopically rather similar have been found: SNe~1999cq, 2000er and 2002ao
\citep[discovered by][respectively]{mod99,cha00,mar02}. \citet{fol07} first proposed that 2006jc-like
events may constitute a distinct class of core-collapse SNe exploding in a He-rich CSM.  
SN~2005la \citep{puk05} is also a somewhat similar SN, although with some unique properties. 
One might reasonably consider it as intermediate  
between a type Ibn and a type IIn event \citep[see][]{pasto07b}. 

\begin{figure*}
\includegraphics[width=8.4cm]{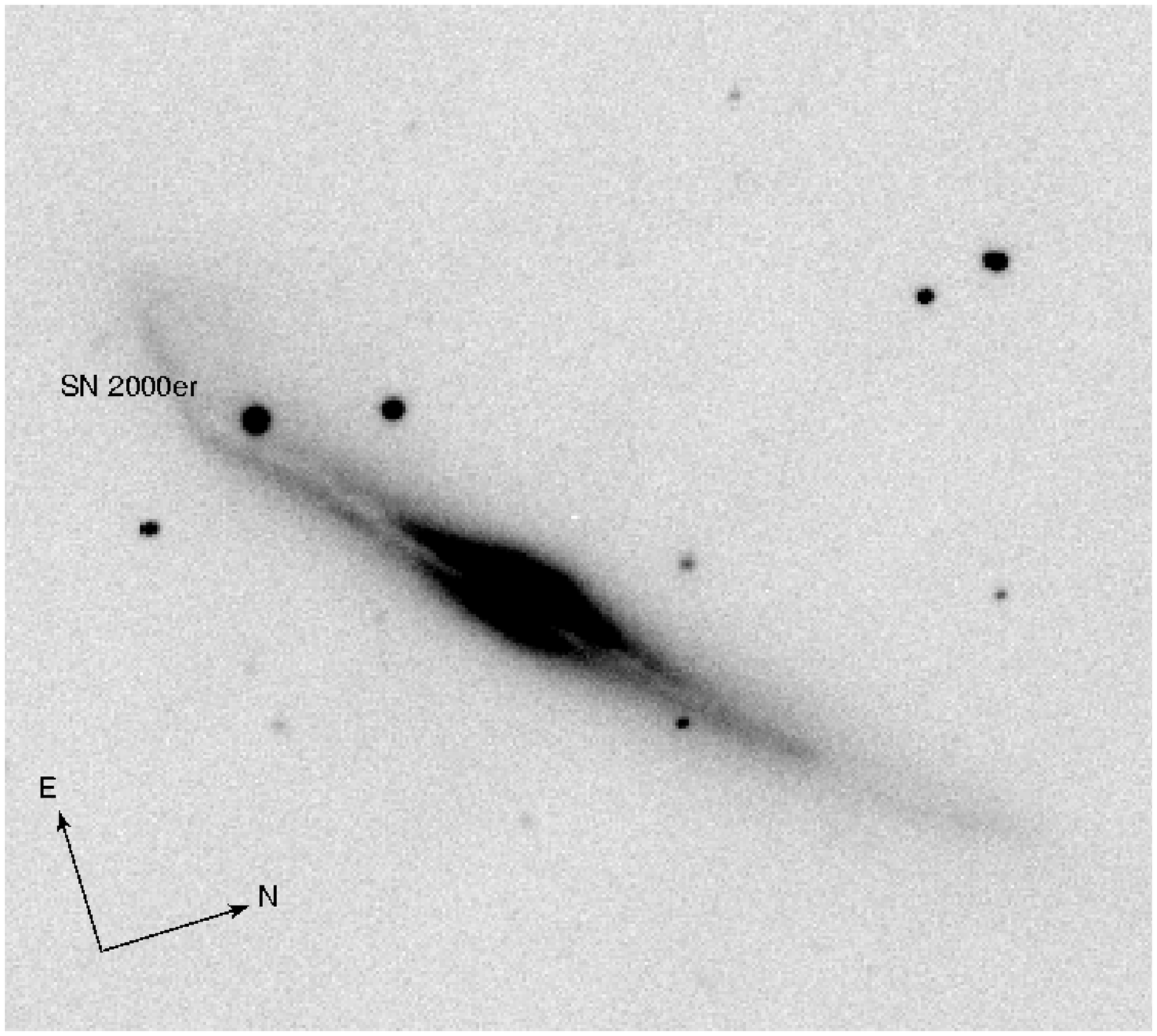}
\includegraphics[width=8.4cm]{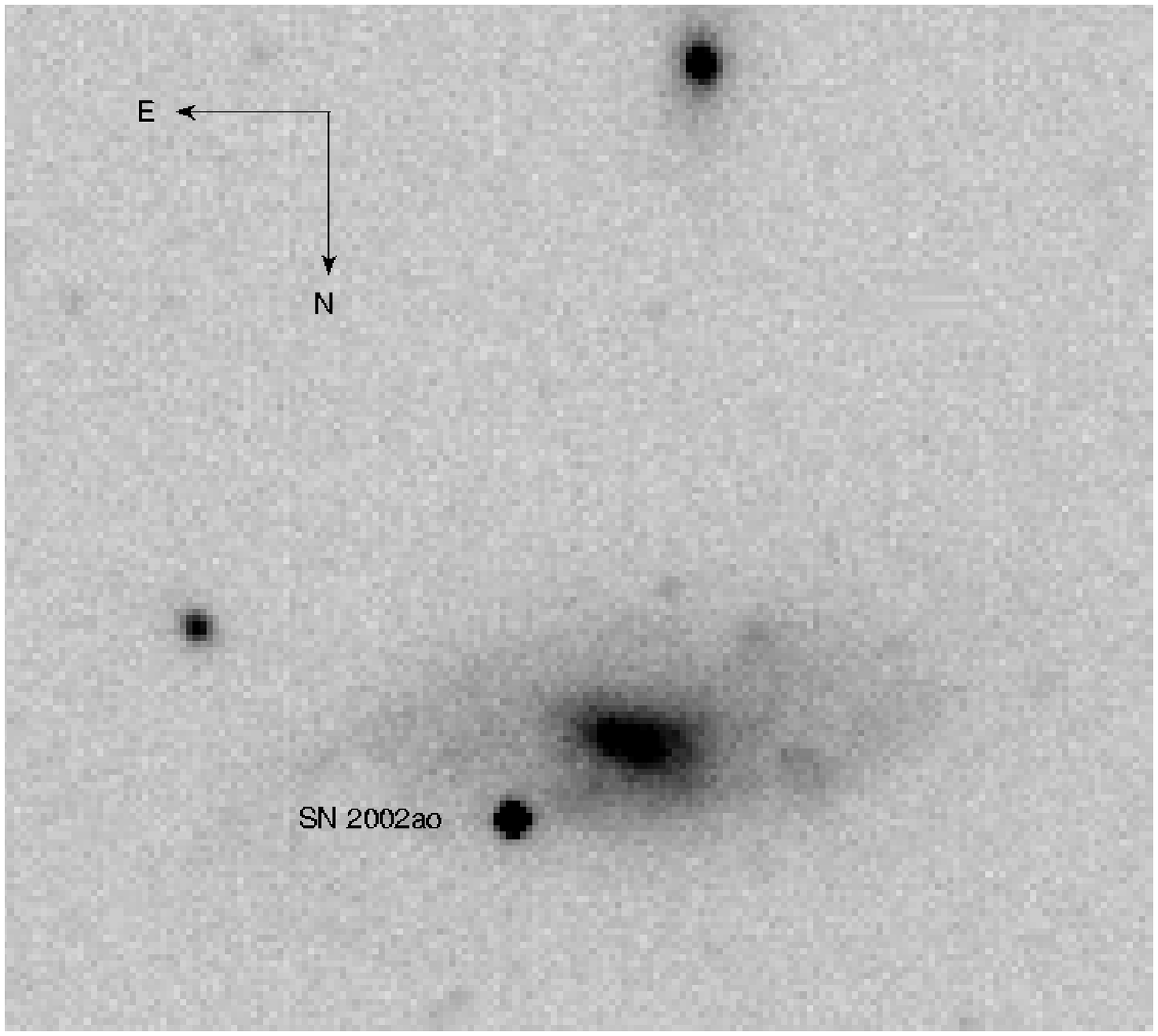}
\caption{R-band images of SN 2000er (left) and SN 2002ao (right). 
They were obtained 
on November 26, 2000 using the ESO 3.6m Telescope in La Silla (Chile) 
and  on February 7, 2002 using the IAC 80-cm, in La Palma (Canary Islands, Spain), respectively. 
\label{fig2}}
\end{figure*}

This article is the first of a series of three papers \citep[together with][]{pasto07b,seppo07} in which we will discuss 
some peculiar, transitional events possibly produced by the explosions of very massive stars.
In this work we will focus on the properties of the 4 objects which belong to the type Ibn SN family. 
Some of the data analysed in this paper have already been published \citep{mat00,fol07,pasto07},
others have never been shown before.
For SN 1999cq we do not present new data, but we consider only those of
\citet{mat00}.   
The afore mentioned SN~2005la, which has even more peculiar characteristics, will be analysed separately \citep{pasto07b}.

The paper is organised as follows: in Section \ref{host} the properties of the host galaxies 
of the SN sample are introduced, including reddening and distance estimates. Spectroscopic  and photometric 
observations are presented in Section \ref{sp}
and Section \ref{lc}, respectively. The quasi-bolometric light curve of SN~2006jc is presented in  Section \ref{bolo},
while the parameters derived from the bolometric light curve modelling are discussed in
Section \ref{model}. 
In Section \ref{prog} we examine two different plausible scenarios for the progenitors of SN~2006jc and similar events,
while in Section \ref{rate} we estimate the frequency of type Ibn events. Finally, a short summary is 
reported in Section \ref{end}.

\section{SNe Ibn and their host galaxies} \label{host}

\begin{table*}
\footnotesize
\caption{Basic information on our SN sample. Distances (d, column 7) and distance moduli ($\mu$, column 8) have been derived from the recession velocities
of LEDA corrected for Local Group infall into the Virgo Cluster, and assuming H$_0$ = 71 km s$^{-1}$ Mpc$^{-1}$. 
In column 6 the integrated metallicity of the host galaxies (see text) is reported. 
In column 9 the Galactic reddening from \citet{sch98} and in column 10 the adopted (total) reddening are provided, while in column 11
the Julian Date of the light curve peak is indicated.
The extinction law of \protect\cite{car89}, with R$_V$=3.1, is adopted throughout this paper. \label{tab1}}
\begin{tabular}{ccccccccccccc}
\hline\hline
SN & $\alpha$ & $\delta$ & Galaxy & Type & 12+log(O/H)& d &  $\mu$ & E$_G$(B-V) & E(B-V) & JD(max) & Reference\\
&&&&& (dex)&(Mpc) & (mag)&(mag)&(mag)& +2400000 &  \\
\hline
1999cq & $18^{h}32^{m}07\fs10$ & $+37\degr36\arcmin44\farcs3$ & UGC 11268 & Sbc & 9.3 & 114 &  35.29 & 0.054 & 0.15 & 51348$\pm$4 & $\circledcirc$ \\
2000er & $02^{h}24^{m}32\fs54$ & $-58\degr26\arcmin18\farcs0$ & PGC 9132 & Sab &  9.2 & 128 & 35.53  & 0.033 & 0.11 & 51869$\pm$8 & $\ddag$ \\
2002ao & $14^{h}29^{m}35\fs74$ & $-00\degr00\arcmin55\farcs8$ & UGC 9299 & SABc & 8.6 & 23 & 31.79         & 0.043 & 0.25 & 52283$\pm$10 & $\ddag$, $\diamondsuit$ \\
2006jc & $09^{h}17^{m}20\fs78$ & $+41\degr54\arcmin32\farcs7$ & UGC 4904 & SBbc & 8.5 & 26 & 32.04         & 0.020 & 0.02 & 54008$\pm$15 & $\ddag$, $\diamondsuit$, $\sharp$  \\
\hline
\end{tabular}

$\circledcirc$ \protect\cite{mat00};
$\ddag$ this paper;
$\diamondsuit$ \protect\cite{fol07};
$\sharp$ \protect\cite{pasto07}.\\
\end{table*}

All four objects forming the sample of SNe Ibn analysed in this paper exploded in spiral galaxies,
but showing different characteristics. While the galaxies hosting SNe 1999cq and 2000er are luminous spirals,
both SNe~2002ao and 2006jc are hosted by similarly blue, under-luminous 
(about M$_B \sim$ -16.0 and -18.3 for UGC 9299 and UGC 4904, respectively), barred spiral galaxies.

Using a luminosity-metallicity relation \citep{tre04}, \citet{pri07} find that the host 
galaxies of SNe~2002ao and 2006jc are rather metal-poor. 
On their scale \citep[they adopt the Solar oxygen abundance of 12+log(O/H) = 8.86\,dex of][]{del06}, the
oxygen abundance of these faint galaxies is  12+log(O/H) $\sim$ 8.5\,dex. However, we should bear in mind
that the absolute values of chemical abundances based on nebular lines vary
by as much as 0.5\,dex depending on the calibration \citep{2007ApJ...656..186B}, 
hence when using the \citet{pri07} values we should relate them to their
likely Milky Way solar-type abundances. If we assume an absolute magnitude of 
$-21$ for the Milky Way their ``solar abundance'' would be  
9.0 $\pm$ 0.2\,dex. This would suggest the host galaxies of SNe~2002ao and 2006jc
have subsolar metallicity of 0.4Z$_{\odot}$. 

As mentioned above, the galaxies hosting SNe 1999cq \citep{mat00} and 2000er are
significantly different, being both very luminous (M$_B \sim$ -21.7
and -21.4, respectively) spirals.  \citet{tre04} have
employed Sloan Digital Sky Survey imaging and
spectroscopy of a relatively nearby (0.005 $<$ z $<$ 0.25) large
sample of star-forming galaxies (about 53000) and produced 
an empirical fit between galaxy absolute magnitude and metallicity. 
Again, despite their absolute scale is still uncertain due to difficulties
in calibrating the strong nebular lines \citep{2007ApJ...656..186B}, we will 
adopt this and note the differential abundance between
a typical Milky Way solar-type value of 9.0\,dex 
and the integrated galaxy abundances. 
The galaxy hosting SN 1999cq, UGC 11268, is of Sbc type, with a B-band absolute
magnitude of -21.7 (according to the LEDA\footnotemark
database). \footnotetext{http://leda.univ-lyon1.fr/; \protect\citep{pat03}} The
luminosity-metallicity relation of \citet{tre04} for this magnitude
gives an integrated abundance of 12+log(O/H) $\approx$ 9.3 $\pm$ 0.2 (dex),
significantly higher than those derived for the host galaxies of
SNe~2006jc and 2002ao. The metallicity measurements using the relation
from \citet{tre04} might lead to a slight overestimate of the true
 abundances at the position of SN1999cq because of their selection criteria
\citep[see Section 7 of][]{tre04}, and galactocentric position.
However as SN~1999cq lies in the central part of the host galaxy
(4.3$\arcsec$ off-centre, i.e. with a de-projected distance of 2.4 kpc
from the nucleus) we would expect the  metallicity at the SN location 
not to deviate too much from the global value quoted.

 \begin{table*}
\footnotesize
\caption{Main information on the new spectra of SNe 2006jc, 2000er and 2002ao presented 
in this paper. \label{tab2}}
\begin{tabular}{cccccc}
\hline\hline
Date & JD & Phase (days)& Instrumental configuration & Range (\AA) & Resolution (\AA) \\
\hline
\multicolumn{6}{c}{SN 2006jc} \\
\hline
Dec 20, 2006 & 2454089.60 & 81.6 & Ekar 1.82m + AFOSC + gr.4 & 3650-7800 & 24\\
Dec 26, 2006 & 2454095.74 & 87.7 & WHT 4.2m + ISIS + R300B+R158R & 3200-9550 & 7+10 \\
Feb 08, 2007 & 2454139.64 & 131.6 & WHT 4.2m + ISIS + R300B+R158R & 3050-9080 & 7+10 \\
Mar 22-23-24, 2007$^{1}$ & 2454182.58 & 174.6 & VLT UT1+ FORS2 + gr.300I + OG590 & 5900--11100 & 10 \\
\hline
\multicolumn{6}{c}{SN 2000er} \\
\hline
Nov 25, 2000 & 2451873.69  & 4.7 & ESO 3.6m + EFOSC2 + gr.5 & 5170-9270 & 16 \\
Nov 26, 2000 & 2451874.68  & 5.7 & ESO 3.6m + EFOSC2 + gr.11 & 3350-7230 & 16 \\
Nov 27, 2000 & 2451875.72  & 6.7 & ESO 3.6m + EFOSC2 + gr.13 & 3670-9290 & 20 \\
Nov 29, 2000 & 2451877.67  & 8.7 & ESO 3.6m + EFOSC2 + gr.13 & 3670-9290 & 20 \\
Nov 30, 2000 & 2451878.70  & 9.7 &  ESO 3.6m + EFOSC2 + gr.13 & 3670-9290 & 20 \\
\hline
\multicolumn{6}{c}{SN 2002ao} \\
\hline
Jan 30, 2002 & 2452304.62   & 21.6 & Wise 1m + FOSC + gm.600 & 4150-7800 & 18.5 \\ 
Jan 31, 2002$^{2}$ & 2452306.24   & 23.2 & Gunma 0.65m + GCS + 300gr./mm & 3810-7650 & 12 \\ 
Feb 01, 2002 & 2452306.61   & 23.6 & Wise 1m + FOSC + gm.600 & 4150-7800 & 18.5 \\  
Feb 05, 2002 & 2452311.26   & 28.3  & AAT + RGO + gm.300b & 3280-9290 & 6.5 \\  
Feb 08, 2002 & 2452313.60   & 30.6  & Wise 1m + FOSC + gm.600 & 4150-7800 & 18.5 \\ 
\hline
\end{tabular}


$^{1}$ The final spectrum shown in this paper is obtained averaging 
6 individual VLT spectra with 2820 seconds of exposure time each, and obtained during three
subsequent nights: March 22, 23 and 24. The mean Julian Day (JD) and the phase are 
computed averaging the JDs of the individual observations.

$^{2}$ Not shown in Fig. \ref{fig4} because of its low signal-to-noise.

\end{table*}

SN~2000er occurred in the edge-on peculiar/interacting Sab ESO 115-G9 galaxy,
which has a B-band absolute magnitude of -21.4 (LEDA). The Tremonti et al. (2004) relation gives
an integrated oxygen abundance 12+log(O/H) $\approx$ 9.2 $\pm$ 0.2 (dex). 
Even though early-type spirals have shallow  abundance gradients, in this case the effect may not
be negligible at the supernova distance. Since the host galaxy inclination is
close to 90 degrees, adopting a distance of ESO 115-G9 of about 130 Mpc, the de-projected distance of the SN 
is about 30$\pm$10 kpc. 
If we assume that the abundance gradient is similar to the mean value, about -0.02 (dex) kpc$^{-1}$ 
obtained by \citet{dut99} for this galaxy type, 
the expected abundance at the SN location will be $\sim$8.9$\pm$0.2 dex, which is around solar. 
                                                                              
Basic information on the four SNe and their host galaxies (taken from LEDA)
is reported in Table \ref{tab1}.
Two R-band images of SN~2006jc at late phases are shown in Fig. \ref{fig1}, and R-band images of SNe 2000er and 2002ao
in Fig. \ref{fig2}. An image of the host of SN 1999cq can be found in the Lick Observatory Supernova Search 
web pages.\footnotemark
\footnotetext{\sl http://astro.berkeley.edu/$\sim$bait/1999/sn99cq.html}
The distances of the 4 host galaxies were determined using the  
recession velocities corrected for the effect of the Local Group infall into the Virgo Cluster and a value of 
H$_0$ = 71 km s$^{-1}$ Mpc$^{-1}$. 
The adopted distance moduli are reported in Table \ref{tab1}.

The unusual blue colour observed in SN~2006jc and its relatively peripheral location within the host galaxy 
suggest that this SN suffered very little interstellar extinction. 
Hereafter we will adopt for SN~2006jc a total reddening of
E(B-V) = 0.02 \citep[equal to the Galactic component,][]{sch98}, in agreement with the values adopted by
\citet{pasto07,fol07,smi07}.
By contrast, there is indication of some 
additional extinction inside the parent galaxies for the other three SNe. 

The low-resolution spectra of SN~2000er show evidence of narrow interstellar Na ID at the rest
wavelength of the host galaxy,  indicating that some dust in the parent galaxy is extinguishing the SN light.
From the equivalent width of the doublet (EW = 0.48\AA) obtained averaging the measurements from 5 different
spectra, and using the \citet{tura03} relation, we obtain E(B-V) = 0.08. Taking into account the 
Galaxy contribution from \citet{sch98} we derive a total E(B-V) = 0.11. Similarly, for SN 1999cq an upper limit of 
E(B-V) $\leq$ 0.45 was obtained by \citet{mat00} using again the EW of the Na ID interstellar 
line. However, applying other techniques, they found this limit to be probably not stringent, and
a more reliable upper limit was estimated to be E(B-V) $\leq$ 0.25, although an even lower, 
possibly negligible extinction was not excluded. 
Hereafter we will adopt $E(B-V) = 0.15$ as a 
best value for SN~1999cq,  most of which is due to the host galaxy. 
Finally, the spectra of SN 2002ao have modest signal to noise and the detection of narrow interstellar 
lines is not trivial. We will adopt a total reddening
E(B-V) = 0.25, as obtained from a comparison with the colours of SN~2006jc at similar phases \citep[Section
\ref{sp2}, but also Fig. 3 in][]{pasto07b}. 

The values of the total interstellar reddening adopted for the 4 SNe of our sample are reported in Table \ref{tab1}
along with the epochs of the light curve maxima, as estimated using spectroscopic
and photometric comparison criteria (see Section \ref{sp2} for more details). Hereafter, throughout the paper,
the phases computed from the epochs of the maxima will be used.

\section{Spectroscopic Observations} \label{sp}

\begin{figure*}
\includegraphics[width=13cm]{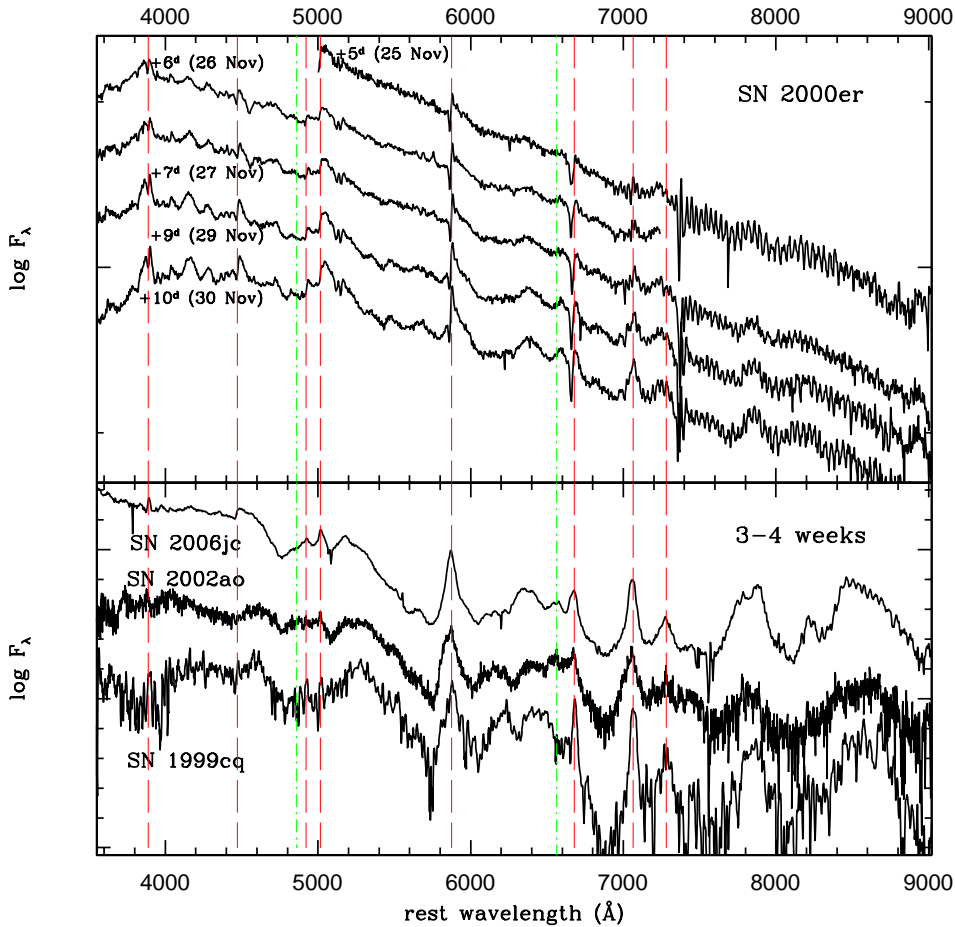}
\caption{Top: spectroscopic evolution of SN~2000er soon after the explosion.
Bottom: comparison with spectra of other SNe Ibn, observed at later phases. Vertical red
dashed lines mark the position of the main He I features, while the dot-dashed
green lines reveal the rest wavelength position of H$\alpha$ and H$\beta$. 
All spectra have been dereddened as indicated in Table \ref{tab1}. 
Note the broad OI $\lambda$7774 feature beginning to develop in the +7 days spectrum, and
becoming stronger at +10 days. The v$_{FWHM}$ of this feature is about 5300 $\pm$ 1100 km s$^{-1}$.
\label{fig3}}
\end{figure*}

The 4 SNe discussed in this paper were initially classified as peculiar type Ib/c SNe
\citep{fil99,mau00a,clo00,den00,gal02a,kin02,gal02b,fil02,cro06,fes06,ben06a,mod06}.
In this section we present some of these classification spectra plus a few unpublished 
spectra of SNe ~2006jc, 2000er and 2002ao (details are listed in Table \ref{tab2}). 
These spectra are analysed 
along with those shown in previous publications \citep{mat00,pasto07,fol07}, and the  
implications are discussed below.

\subsection{Early Spectra} \label{sp1}

\begin{figure*}
\includegraphics[width=13cm]{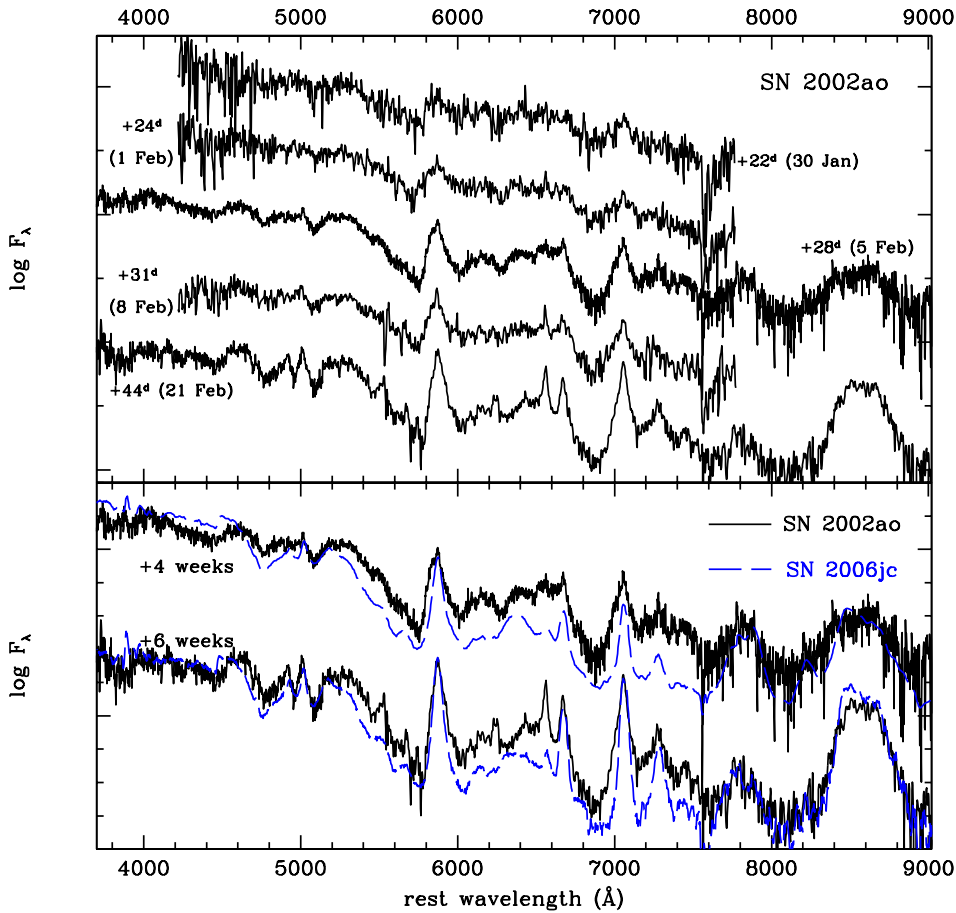}
\caption{Top: spectroscopic evolution of SN~2002ao. The 21st February spectrum
from \protect\cite{fol07} has been included. Bottom: comparison of spectra of SN~2002ao and SN~2006jc \protect\citep{pasto07,fol07}
at about 4 weeks and 6 weeks past maximum. The spectra of SN~2002ao have been 
dereddened by E(B-V)=0.25 magnitudes, while those of SN~2006jc by  E(B-V)=0.02 magnitudes.\label{fig4}}
\end{figure*}

The only type Ibn SN for which early phase spectra are available is SN~2000er. 
Those spectra are shown in Fig. \ref{fig3} (top panel) and
a comparison with later spectra of other SNe Ibn  is shown in the bottom
panel \citep{pasto07,fol07,mat00}. 

The spectra of SN~2000er are quite blue,
in agreement with its adopted young age at the time of the observations (see Table \ref{tab2}). 
The first spectrum (November 25) is almost featureless:
only the narrow P-Cygni lines of He I are clearly visible. 
The subsequent spectra start to develop more prominent He I emission components, while there is no evidence for the H Balmer lines. 
The prominent  He I lines show narrow P-Cygni profiles. This is a significant difference 
compared to the spectra of other SN Ibn events presented so far \citep{mat00,fol07,pasto07}. 
These narrow components are indicative of the presence of a slowly expanding 
($\sim$800-900 km s$^{-1}$) CSM around the SN. The same He I lines show however also broader wings, with
 full-width-half-maximum (FWHM) velocity v$_{FWHM} \sim$5000 km s$^{-1}$, suggesting that some helium is 
also present in the 
fast-moving SN ejecta. 
A broad OI $\lambda$7774 feature is detected with a v$_{FWHM}$
consistent with that of the broad He I line wings. The broad Ca II IR feature has not yet developed. 
Although  the spectra of SN~2000er are from an early phase, they show most of
the features detected in the spectra of other SNe Ibn (Fig. \ref{fig3}). 
 
We notice that SN 1999cq was discovered very young \citep{mat00}, and of all the SNe Ibn it is very likely that 
this is the one discovered closest to the explosion epoch. 
However, the first spectrum by \cite{mat00} was not taken until a few weeks after the discovery announcement,
hence an opportunity for very early spectral coverage was missed for this object.

\subsection{Intermediate-Age Spectra and Phase Calibration} \label{sp2}

Using the spectra available for the 4 SNe, along with the photometric data (see Section \ref{lc}), 
we have attempted a consistent estimate of the phase of all the objects of our type Ibn SN sample.
The almost featureless spectra of SN~2000er and their rapid evolution (Fig. \ref{fig3}) are indicative 
that the SN was probably discovered within about one week from the explosion. Similarly, the photometric data of
SN~1999cq, and in particular the stringent pre-discovery limit, suggest that 
this object was also discovered very young \citep[see][and Section \ref{lc}]{mat00}. 
By contrast, the presence of more prominent
spectral features in the SN 2006jc spectra indicates that this object was discovered at a 
slightly later phase, a couple of weeks after the explosion or $\sim$10 days after maximum light. 
As a further consistency check on the phase calibration,
we have verified that the SN 1999cq spectrum presented in \citet{mat00} closely resembles the spectrum 
of SN~2006jc obtained on October 24th, 2006.

\begin{figure*}
\includegraphics[width=13cm]{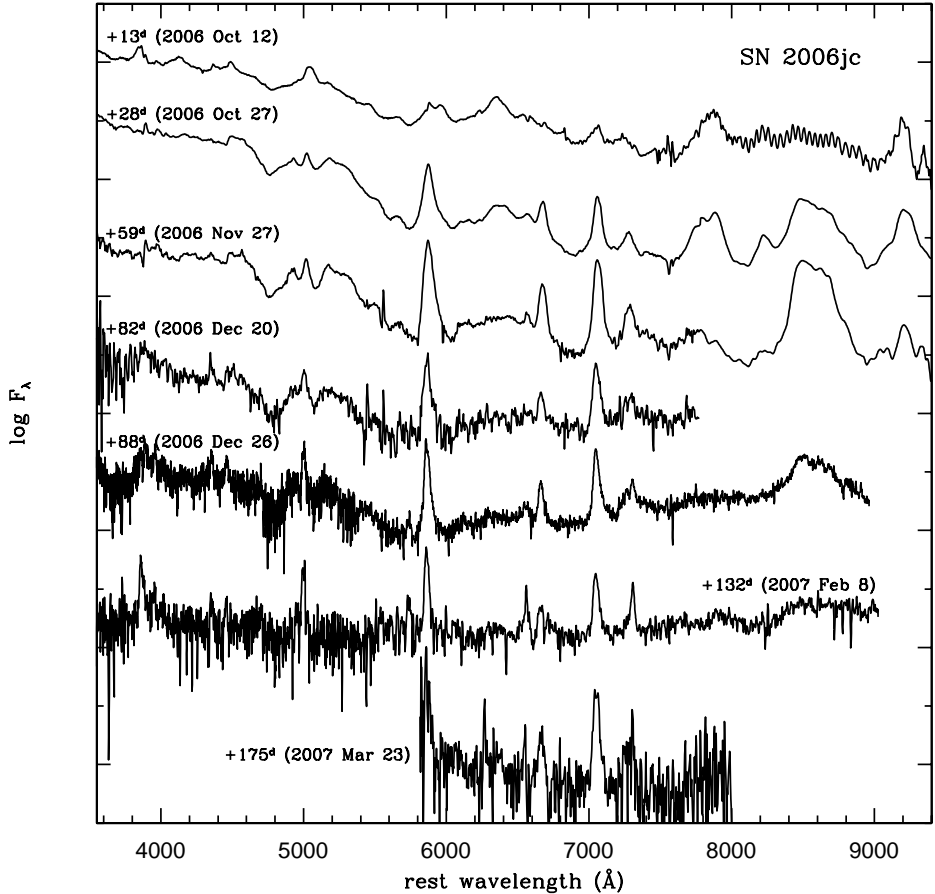}
\caption{Spectroscopic evolution of SN~2006jc. Some early time spectra from \protect\cite{pasto07}
plus unpublished late-time spectra are shown. \label{fig5}}
\end{figure*}

Dating SN~2002ao is less straightforward because this object was discovered late and it was not intensively
monitored. 
Some spectra  of SN~2002ao, obtained when the SN was already quite evolved compared to SN~2000er,
are shown in Fig. \ref{fig4}. 
A unique and remarkable feature of the spectrum at +6 weeks is that the He I
lines show a double-component profile, with a narrow component (900-1500 km s$^{-1}$) on top of
a broader one (4000-5000 km s$^{-1}$).
This broader component was indeed not evident in the spectra of SN~2006jc (see e.g. the unblended 
He I 7065\AA~ feature in Fig. \ref{fig4}, bottom panel).
This indicates that, although most He was lost by the progenitor of SN~2002ao during the WR phase, 
some He was still present in the star's envelope and ejected at the time of the SN explosion (analogous to 
that observed in SN~2000er). Apart from this difference, the spectra of SN~2002ao are rather similar to those of 
SN~2006jc obtained 1-2 months past-maximum. The phase of these spectra was therefore calibrated with reference 
to SN~2006jc, using as a reference a best-fit comparison code \citep[{\bf Passpartoo},][]{avik07}.
We gave more weight to the bluest (below $\sim$5500 \AA) and
the reddest (beyond about 7200 \AA) regions of the spectrum than to the intermediate wavelength region (5500-7200
\AA) as these were less contaminated by the prominent narrow circumstellar lines. 
In particular, we note that the highest signal-to-noise (S/N) spectra of SN~2002ao obtained on February 5th and 
February 21st \citep[][see also Fig. \ref{fig4}]{fol07} provide a good fit (in terms of line velocity and relative strengths of the 
broad lines) to the spectra of SN~2006jc at phases +24 and +41 days respectively \citep{pasto07}.

\begin{figure*}
\includegraphics[width=11.5cm]{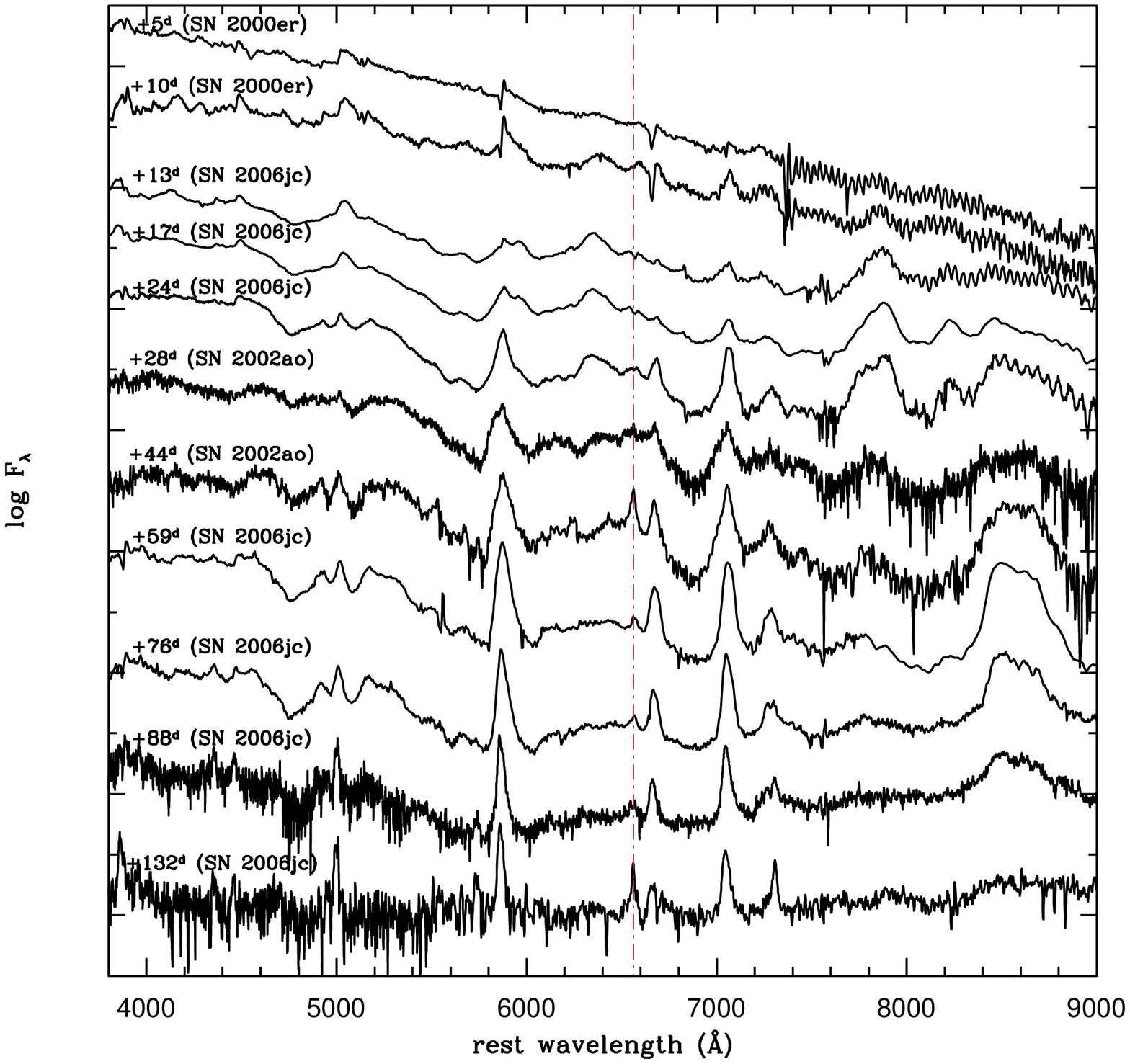}
\caption{Spectroscopic evolution for a representative sample of SNe Ibn. The spectra are reddening-corrected. 
The +44$^d$ spectrum of SN~2002ao is from \protect\cite{fol07}; those of SN~2006jc from 
+13$^d$ to +76$^d$ are from \protect\cite{pasto07}. The vertical dot-dashed line marks the H$\alpha$ rest wavelength. \label{fig6}}
\end{figure*}

\subsection{Nebular Spectra} \label{sp3}
Contrary to other type Ibn SNe for which only few sparse observations exist,
SN~2006jc was extensively monitored during the first $\sim$70 days after discovery \citep{pasto07,fol07}. 
This makes SN~2006jc the best observed object of this class. The existing spectral database of
SN~2006jc covers its evolution starting from about 10 days after the inferred maximum light (see 
Table \ref{tab1} and Section \ref{sp2}).
Some early spectra from \citet{pasto07} and new late-time spectra of SN~2006jc are shown in Fig. \ref{fig5}.
The VLT spectrum was obtained by merging a sequence of 
spectra obtained between March 22nd and 24th, 2007 (with an integrated exposure time of 
more than 6 hours). This is, to our knowledge, the latest optical spectrum existing for an object of this class. 

The last 2 spectra (February and March, 2007) are completely different from those obtained
in December, 2006 (see Fig. \ref{fig5}). They do not show  the strong blue continuum characterizing
the spectra of SN~2006jc at earlier phases or the flux excess at the redder wavelengths discussed by \citet{smi07}. 
Moreover there is no clear evidence of the broad emission lines characterising SNe Ib/c at similar phases. 
Instead, prominent and 
narrow (v$_{FWHM} \approx$ 1800-1900 km s$^{-1}$) He I lines can be identified and still 
dominate the spectrum about 6 months after the explosion. A
 further difference to earlier spectra of SN 2006jc lies in the strength of H$\alpha$, which is
 almost comparable with the 6678\AA~ He I line  in the February spectrum (phase $\sim$ +132 days; see
Fig. \ref{fig5}), although even narrower (v$_{FWHM} \sim$ 1000 km s$^{-1}$). 
This indicates the presence of an outer, H-rich circumstellar shell beyond the He I layer 
which is likely ionised by photons from ejecta-CSM interaction or by high-energy photons propagated through  the He-rich region.

\subsection{Overall Evolution} \label{sp4}

The spectra collected for our SN Ibn sample allow us to illustrate the spectroscopic evolution for this 
SN type (see Fig. \ref{fig6}). The phases are calibrated with reference to the epochs of maximum reported in Table \ref{tab1}.

\begin{figure}
\includegraphics[width=8.5cm]{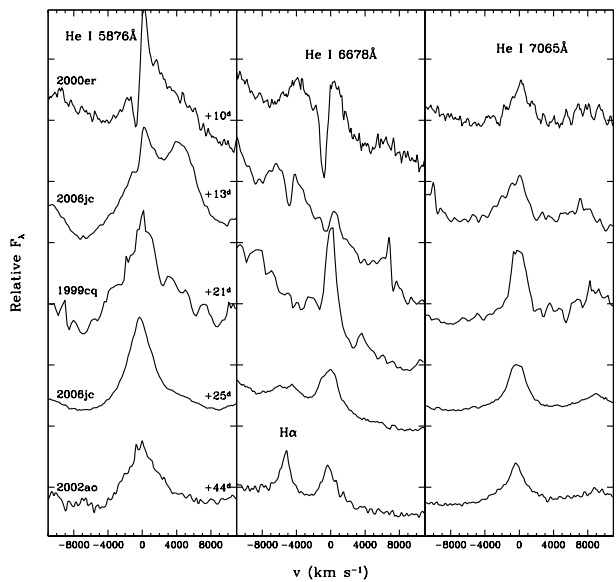}
\caption{Profiles of some selected He I lines for a representative sample of spectra of SNe Ibn. The spectra are reddening-corrected. 
The +21$^d$ spectrum of SN~1999cq is from  \protect\cite{mat00}, that of SN~2002ao at
+44$^d$ is from \protect\cite{fol07}, while those of SN~2006jc from 
+13$^d$ to +25$^d$ are from \protect\cite{pasto07}. \label{fig6b}}
\end{figure}

Despite a considerable uncertainty in the explosion epochs and in the interstellar reddening estimates, 
there appears to be a remarkable homogeneity in the spectroscopic properties among the objects. 
This is also evident from a check of the profiles of the narrow He I lines in our SN sample. 
The evolution of the line profiles of the features at 5876\AA, 6678\AA~and 7065\AA~is shown in Fig. \ref{fig6b} 
and has been obtained making use of the spectra of 4 different SNe Ibn.
This homogeneity in the spectral properties would be quite surprising if the ejecta were strongly interacting with a pre-existing 
CSM, because the density, geometry and composition of the CSM would be  expected to affect
significantly the evolution of SNe Ibn.
One possibility is that the homogeneity in the observed properties of SNe Ibn may indicate that the
ejecta-CSM interaction is playing a minor role in the evolution of these SNe.

The most significant difference among SNe Ibn lies 
in the strength of the narrow H spectral lines (Fig. \ref{fig6}, where the position of H$\alpha$ is marked by a dot-dashed vertical line, see
also Fig. \ref{fig6b}, central panel). 
This likely indicates some variation in the abundance of H in the 
innermost CSM region. To a lesser degree there are differences in the widths and profiles of the circumstellar He I lines.

\section{Light Curves}  \label{lc}

New photometric data of SNe 2000er, 2002ao and 2006jc are reported in Table \ref{tab3}. The photometric measurements in the reduced images
(i.e. after overscan, bias and flat field corrections) have been performed using standard PSF-fitting procedures
in the cases of SNe~2000er and 2006jc, since reliable galaxy templates were not available. Conversely, 
for SN~2002ao we have subtracted the host galaxy contribution
using images of the SN field obtained  in April 2007, using the 1.82m Copernico Telescope 
of Mt. Ekar (Asiago, Italy). Photometric zeropoints were then computed 
using standard fields from the \citet{lan90} catalogue observed during some photometric nights, 
and then the SN magnitudes were calibrated after comparison with the magnitudes of reference stars in 
the fields of the three SNe \citep[for more details on the method see e.g.][]{pasto06}. The local sequence calibrated 
for SN~2006jc is the same as adopted in \citet{pasto07}.

\begin{table*}
\footnotesize
\caption{Additional, unpublished late time photometry of SN 2006jc, plus new 
photometry for SNe 2000er and 2002ao. \label{tab3}}
\begin{tabular}{ccccccccc}
\hline\hline
Date & JD & Phase & U & B & V  & R & I & Instrument \\
\hline
\multicolumn{9}{c}{SN 2006jc} \\
\hline
Dec 20, 2006 & 2454089.65 & 81.7 &              & 19.53 (0.09) & 19.53 (0.07) & 19.16 (0.08) & 18.16 (0.06) & Mt.Ekar-1.82m+AFOSC \\
Dec 22, 2006 & 2454091.57 & 83.6 & 18.95 (0.05) & 20.03 (0.06) & 19.98 (0.06) & 19.69 (0.06) & 18.44 (0.03) & LT-2.0m+RatCam \\
Dec 30, 2006 & 2454099.51 & 91.5 & 19.67 (0.32) &              & 20.74 (0.21) & 20.27 (0.17) & 19.05 (0.07) & LT-2.0m+RatCam \\
Jan 08, 2007 & 2454109.51 & 101.5 & $>$20.21  & 21.62 (0.22) & 21.64 (0.34) & 21.15 (0.22) & 19.90 (0.12) & LT-2.0m+RatCam \\ 
Jan 14, 2007 & 2454114.70 & 106.7 &              & 22.11 (0.22) & 22.04 (0.38) & 21.60 (0.28) & 20.12 (0.25) & LT-2.0m+RatCam \\ 
Jan 17, 2007 & 2454117.70 & 109.7 &              & 22.57 (0.13) & 22.59 (0.26) & 21.85 (0.18) & 20.30 (0.10) & LT-2.0m+RatCam \\ 
Feb 06, 2007 & 2454138.44 & 130.4 &              & $>$22.71  & $>$22.70  & 22.93 (0.38) & 21.17 (0.41) & LT-2.0m+RatCam \\ 
Feb 07, 2007 & 2454139.45 & 131.5 &              &              &              & 23.05 (0.21) & 21.33 (0.14) & WHT-4.2m+Aux Port Imager  \\
Feb 08, 2007 & 2454139.78 & 131.8 &              &              & 23.59 (0.38) &              &              & WHT-4.2m+Aux Port Imager  \\  
Feb 11, 2007 & 2454143.38 & 135.4 &              & $>$23.29  & $>$23.10  & 23.46 (0.41) & 21.64 (0.28) & LT-2.0m+RatCam \\ 
Mar 11, 2007 & 2454171.41 & 163.4 &              &              & 24.27 (0.14) & 24.05 (0.25) & 22.96 (0.10) & WHT-4.2m+Aux Port Imager  \\       
Mar 22, 2007 & 2454181.57 & 173.6 &              &              &              & 24.37 (0.45) & 23.30 (0.68) & VLT-UT2+FORS1  \\       
\hline
\multicolumn{9}{c}{SN 2000er$^{1}$} \\
\hline
Nov 25, 2000 & 2451873.68 & 4.7  & &  &  & 16.29 (0.01) & & ESO-3.6m+EFOSC2 \\
Nov 25, 2000 & 2451873.69 & 4.7  & &  &  & 16.31 (0.01) & & ESO-3.6m+EFOSC2 \\
Nov 26, 2000 & 2451874.67 & 5.7  & & 16.73 (0.01) & 16.47 (0.01) & 16.36 (0.01) & & ESO-3.6m+EFOSC2 \\
Nov 27, 2000 & 2451875.70 & 6.7  & &  &  & 16.54 (0.01) & & ESO-3.6m+EFOSC2 \\
Nov 29, 2000 & 2451877.65 & 8.7  & &  &  & 16.65 (0.01) & & ESO-3.6m+EFOSC2 \\
Nov 29, 2000 & 2451877.66 & 8.7  & &  &  & 16.65 (0.01) & & ESO-3.6m+EFOSC2 \\
Nov 30, 2000 & 2451878.68 & 9.7  & &  &  & 16.74 (0.01) & & ESO-3.6m+EFOSC2 \\
Feb 02, 2001 & 2451942.54 & 73.5  & & 23.42 (0.10) & 23.38 (0.20) & 23.32 (0.17) & & ESO-3.6m+EFOSC2 \\
\hline
\multicolumn{9}{c}{SN 2002ao} \\
\hline
Feb 07, 2002 & 2452312.74 & 29.7  & & 16.79 (0.03) & 16.51 (0.02) & 16.21 (0.02) & 16.09 (0.02) & IAC-0.80m+CCD \\
Feb 08, 2002 & 2452313.75 & 30.8  & & 16.88 (0.06) & 16.60 (0.02) & 16.29 (0.02) & 16.21 (0.03) & IAC-0.80m+CCD \\
Feb 10, 2002 & 2452315.66 & 32.7  & & 17.09 (0.05) & 16.80 (0.04) &              & 16.37 (0.03) & IAC-0.80m+CCD \\
Feb 13, 2002 & 2452319.33 & 36.3  & &              &              & 16.91 (0.05) &              & Kiso-1.05m+CCD \\     
Feb 13, 2002 & 2452319.34 & 36.3  & &              &              & 16.93 (0.04) &              & Kiso-1.05m+CCD \\     
Feb 14, 2002 & 2452320.34 & 37.3  & & 17.75 (0.24) & 17.34 (0.25) & 17.25 (0.26) & 16.97 (0.14) & Kiso-1.05m+CCD \\ 
Feb 18, 2002 & 2452324.62 & 41.6  & & 18.09 (0.04) &              & 17.42 (0.03) &              & Wise-1.0m+CCD   \\
Mar 04, 2002 & 2452338.57 & 55.6  & &              & 18.82 (0.41) & 18.88 (0.55) & 18.37 (0.30) &  Wise-1.0m+CCD  \\
Mar 05, 2002 & 2452339.52 & 56.5  & &              & 18.98 (0.31) & 19.04 (0.28) & 18.43 (0.23) &  Wise-1.0m+CCD  \\
Mar 07, 2002 & 2452341.57 & 58.6  & &              & 19.53 (0.21) & 19.17 (0.27) & 18.88 (0.21) &  Wise-1.0m+CCD  \\
Apr 02, 2002 & 2452366.95 & 84.0  & &              & 22.15 (0.41) & 22.03 (0.44) &              & Kuiper-1.55m+CCD\\
Apr 03, 2002 & 2452367.94 & 84.9  & &              & 22.21 (0.44) &              &              & Kuiper-1.55m+CCD\\
Apr 05, 2002 & 2452369.50 & 86.5  & &              & $>$20.78  & $>$20.61  & $>$20.27  &  Wise-1.0m+CCD  \\
Apr 13, 2002 & 2452377.51 & 94.5  & & $>$20.10  & $>$20.64  & $>$19.77  & $>$19.67  &  Wise-1.0m+CCD  \\
Apr 13, 2002 & 2452378.42 & 95.4  & &              & $>$18.49  & $>$18.91  &              &  Wise-1.0m+CCD  \\
Apr 18, 2002 & 2452383.41 & 100.4  & &              & $>$20.61  & $>$20.25  & $>$19.83  &  Wise-1.0m+CCD  \\
May 05, 2002 & 2452399.84 & 116.8  & & $>$22.14  & $>$22.06  & $>$22.19  & $>$22.00  &  Bok-2.3m+CCD \\
Apr 16, 2007 & 2454206.50 & 1923.5 & & $>$22.69  & $>$22.32  & $>$22.08  & $>$22.04  & Mt.Ekar-1.82m+AFOSC \\ 
Apr 19, 2007 & 2454210.49 & 1927.5 & & $>$21.87  & $>$22.13  & $>$21.86  & $>$21.15  & Mt.Ekar-1.82m+AFOSC \\ 
\hline
\end{tabular}

$^{1}$ K-corrected magnitudes of SN 2000er. The K-correction applied to the data of Feb 2, 2001 was 
estimated using a spectrum of SN~2006jc at similar phase (+74 days), 
reddened by E(B-V) = 0.11 and then shifted to the z value of the galaxy hosting SN~2000er.
\end{table*}

SN 2000er was also observed in J, H, and K bands with the SOFI near-IR
imager mounted on ESO NTT at a single epoch, on 2001 Jan. 11. The SOFI images
were reduced using standard IRAF procedures and the photometry was performed
using PSF fitting procedures. The photometric data were calibrated using
2MASS magnitudes. This yielded magnitudes of 20.20 $\pm$ 0.26, 20.37 $\pm$ 0.30, and
19.45 $\pm$ 0.28 for SN 2000er in the J, H, and K bands, respectively. Apart from
SN 2006jc, SN 2000er is the only type Ibn SN which was observed in the NIR domain.

\begin{figure}
\includegraphics[width=9.0cm]{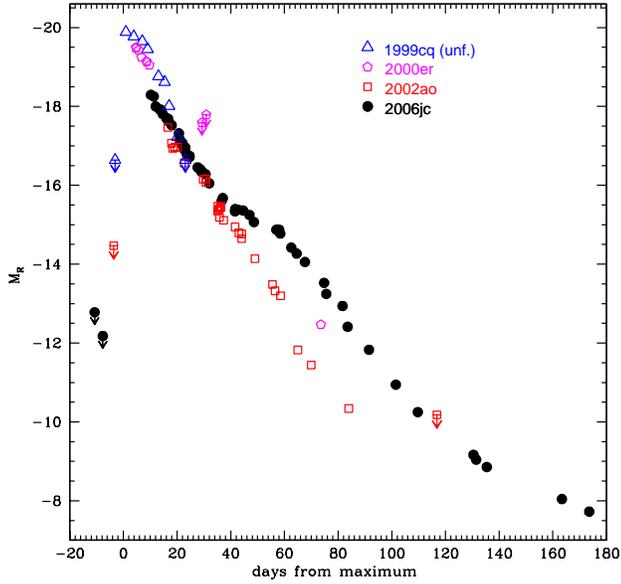}
\caption{R-band absolute light curves for our 2006jc-like SN sample, obtained with
the distance and reddening assumptions of  Table \ref{tab1}. 
The unfiltered observations of SN~1999cq from \protect\citet{mat00} are also shown.  
The most significant detection 
limits are also reported. \label{fig7}}
\end{figure}

In Fig. \ref{fig7} the R-band absolute light curve of SN 2000er is compared with the light curves
of SNe 2006jc \citep{pasto07,fol07}, 1999cq \citep[unfiltered, approximately R band,][]{mat00} and 2002ao \citep{fol07}.
These light curves were computed by adopting the distances and interstellar extinction values reported in
Table \ref{tab1}. Since the host galaxy of SN~2000er has relatively high redshift, a K-correction to the observed magnitudes
was applied. The correction (always within 0.1 magnitudes) was computed making use of the early-time SN spectra plus a late time spectrum of SN~2006jc,
artificially extinguished and redshifted to match the values of E(B-V) and redshift of the host galaxy of SN 2000er. Although also the host galaxy
of SN~1999cq has an high redshift, no K-correction was applied to the data of  \citet{mat00} because of the lack of
spectral coverage. Actually, the redshift effect on the broad-band magnitudes of SN~1999cq might be non-negligible.

SNe~1999cq and 2000er are particularly important because they show the early post-maximum evolution 
of the light curve for representative  objects of this family.
The evolution during the first month after peak is remarkably homogeneous.
A very bright peak magnitude (not far from M$_R \sim$ -20) is probably a common characteristic of SNe Ibn. 
At about 40 days, the light curve of SN~2006jc settles on a slower decline, indicating that the SN
has likely reached the exponential tail. Soon after (between 50 and 60 days), the light
curve becomes again steeper, with the slope being similar to those of 
other SNe Ibn observed at late phases. 
The very rapid luminosity decline may indicate that either a very
small amount of $^{56}$Ni is ejected by the objects of this class, or
that new dust formed in the SN ejecta or in the surrounding
CSM. The formation of new dust is supported by the large infrared
excess observed in SN 2006jc starting from 1 month after the discovery
\citep{ark06,smi07,elisa07,tom07,sak07,seppo07}, a
progressively redder spectral continuum and a clear red-wing attenuation
in the profiles of the He I circumstellar spectral lines \citep{smi07,seppo07}. 
Condensation of dust in the ejecta of a few
core-collapse SNe has been observed at late times \citep[see][and references therein]{meikle07}. 
Moreover, another scenario in which new dust
condensed in a cool dense shell produced by the impact of the SN
ejecta with a dense CSM has previously been proposed to explain the IR
excess observed in the type IIn SN 1998S at epochs later than 300 days
\citep{poz04}. In SN 2006jc it is likely that a similar
mechanism took place but producing new dust at a much earlier epoch
\citep[see][and Section  \ref{bolo} of this paper]{smi07,seppo07}

Unfortunately, so far no SN Ibn  has been observed during the rising phase to maximum. However, some
detection limits were obtained close enough to the first detection and indicate that the rising branch of the light curve 
is extremely fast, similar to what we see in some core-collapse SNe which show evidence of interaction with a CSM 
\citep[e.g SNe 1983K, 1993ad and 1994W,][]{nie85,phi90,pol93,tsv95,sol98,chu04}, but also in more classical non-interacting 
type II SNe \citep[e.g.  SNe~2005cs and 2006bp,][]{pasto07c,qui07}.

\subsection{Quasi-Bolometric Light Curve}  \label{bolo}

\begin{figure}
\includegraphics[width=8.5cm]{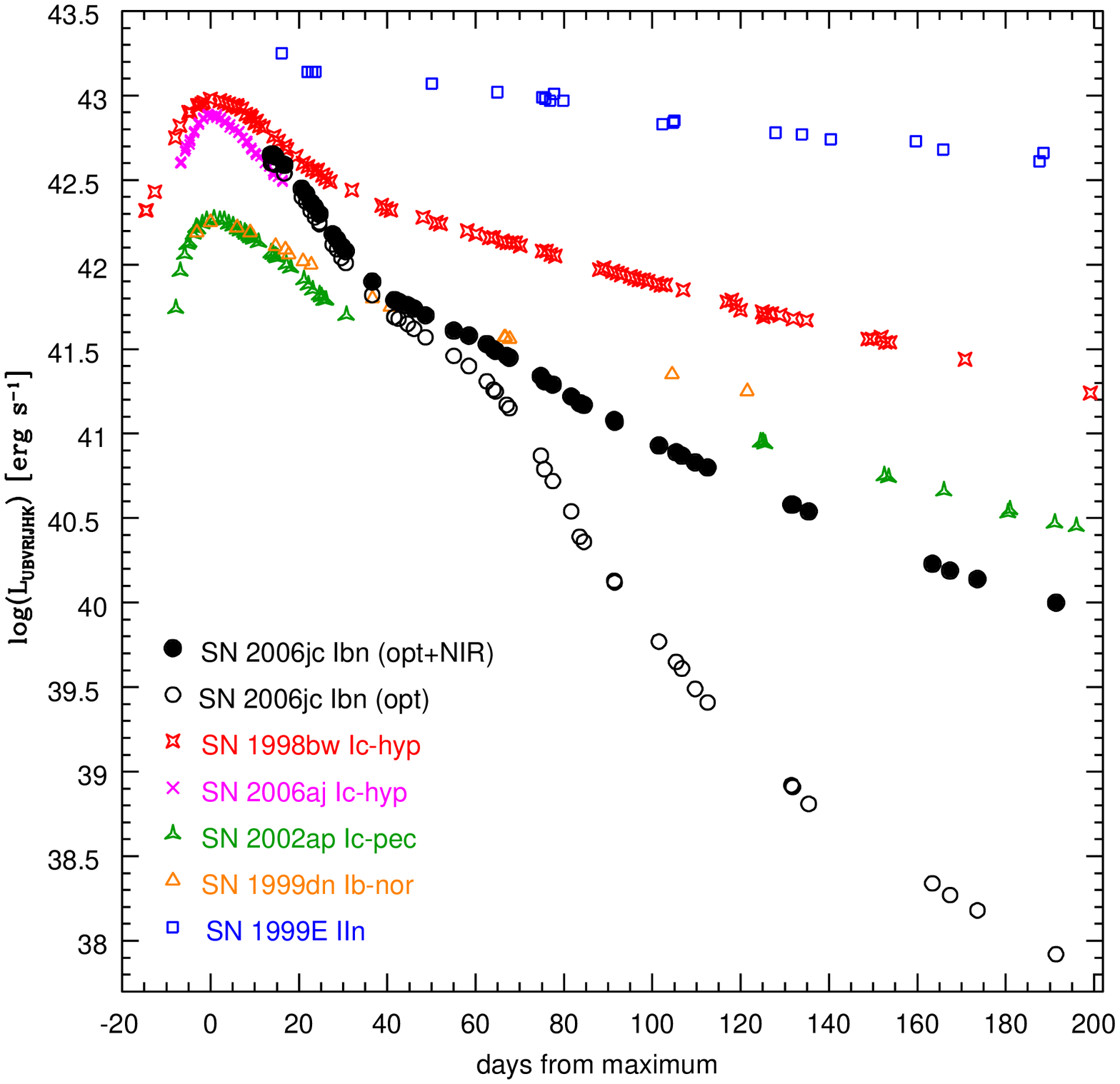}
\caption{Quasi-bolometric (optical+NIR) light curves of  luminous core-collapse SNe: the  
SN Ibn~2006jc (filled circles), the hypernovae SN~1998bw \protect\citep{gal98,mac99,sol00,pat01} and SN~2006aj
\protect\citep{mir06,cob06,pian06,cam06,sol06}; 
the broad-lined but normally luminous type Ic SN~2002ap \protect\citep{yos03,fol03,tom06};
the type Ib SN~1999dn (Benetti et al. in preparation) and the type IIn SN~1999E \protect\citep[][though the core-collapse nature of this 
SN 2002ic-like event is debated]{rig03}. 
The light curve of SN~2006jc obtained integrating the optical bands only is also reported (empty circles), 
showing the dramatic decline of the optical luminosity at $\sim$2 months past-maximum. 
The comparison between the integrated optical and optical plus NIR light curves visualizes that after 
50-60 days most of the SN flux is emitted in the NIR region \citep{seppo07}. The NIR data used to
compute the quasi-bolometric light curve of SN~2006jc are from \citet{ark06}, \citet{koji07} and \citet{seppo07}. \label{fig9}}
\end{figure}

In Fig. \ref{fig9} the quasi-bolometric ({\sl uvoir}) light curve of  the representative type Ibn SN~2006jc (filled circles), is compared 
 with the curves of some well-observed type Ib/c SNe and the 
type IIn SN 1999E \citep[though the core-collapse nature of this SN is disputed due to the similarity to 
SN~2002ic, see e.g.][]{ham03,ben06}. All light curves are computed by integrating the fluxes in the 
optical (UBVRI) and NIR (JHK) bands. 
The {\sl uvoir} curve of  SN~1999E is remarkably flatter and more luminous than that of any other SN in Fig. \ref{fig9}. 
This is because it is mostly powered by the energy output from the interaction between SN ejecta and CSM \citep{rig03}. 
The {\sl uvoir} curve of SN 2006jc is similar to those of non-interacting,
rather luminous core-collapse SNe (Fig. \ref{fig9}), suggesting that the ejecta-CSM interaction
probably plays only a minor role in the luminosity evolution of this SN~Ibn \cite[Section \ref{model}, see also][]{tom07}.

The fast rise, asymmetric, luminous peak and rapid decline 
of the unfiltered light curve of SN~1999cq (see Fig. \ref{fig7}) are reminiscent of some 
bright SNe IIL \citep[e.g. SNe 1979C and 1980K, see][and references therein]{pat94}. 
\citet{bar92} explained the observed properties of the early light curves
of luminous type IIL SNe with a standard core-collapse explosion in which
the optical peak results from the reprocessing of UV photons in a circumstellar
shell generated by a superwind of the SN precursor. 
Alternatively, the early light curve of SNe Ibn might be equally well explained invoking a 
mechanism analogous (though on a shorter time scale) to the H recombination which produces 
the long-duration phase with almost constant 
luminosity (plateau) characterizing SNe IIP. In the case of SNe Ibn, the recombination of the residual He envelope
could be responsible for the asymmetric, luminous light curve peak (see Section \ref{model}). 

In Fig. \ref{fig9} we also show the light curve of SN~2006jc obtained by integrating the flux contribution from the optical
bands only (open circles). Starting from 50-60 days past-maximum, this diverges significantly from the {\sl uvoir} light curve 
(including the NIR contribution, filled circles), suggesting that a significant amount of dust is forming, extinguishing the light at the
optical wavelengths and re-emitting it in the infrared domain \citep[for details, see][]{seppo07}. Since similar steep decays 
in the optical light curves are also observed  in SNe~1999cq and 2002ao \citep[see also][]{fol07}, we could
reasonably argue that the formation of a significant amount of dust might be a common characteristic of most 
type Ibn events. 
Unfortunately, apart from SN 2006jc and only one epoch of NIR
observations of SN 2000er with rather large photometric uncertainties,
no other infrared observations are available for other SNe Ibn to test
for the appearance of an infrared excess contemporaneous with the
rapid optical luminosity decline. This phenomenon is observed at quite
early stages in SN 2006jc and, as already indicated, is probably due
to dust forming in a rapidly cooling shell as a result of the
ejecta-CSM shock interaction. This scenario is discussed extensively
in \citet{seppo07} and \citet{smi07}.

\subsection{Light Curve Modelling}  \label{model}

One way to derive information on the explosion parameters of SNe is the modelling of the bolometric
light curve. Hereafter, we will assume that the interaction between 
ejecta and CSM plays a minor role in the bolometric evolution of SNe Ibn. 
This is somewhat discrepant with some observed spectral features and with the X-ray detection of at least one of them, SN 2006jc
\citep{bro06,imm06,imm07b}, both of which would suggest some interaction is actually occurring. 
However, the X-ray luminosity \citep{imm07b} is rather low compared to the IR luminosity at the same epochs,
and this would be against a strong interaction scenario.
Then, SNe Ibn show fast-evolving quasi-bolometric 
light curves (see Fig. \ref{fig9}), homogeneous spectro-photometric evolution and SN 2006jc was undetected 
at radio wavelengths soon after its discovery. The radio luminosity of SN 2006jc was indeed 
at least 100 times less than that of the broad-line SN Ic 1998bw \citep{sod06}, and at most
comparable with those of normal, non-interacting type Ib/c SNe \citep{ber03}. 
All of this seems inconsistent with a scenario in which the SN ejecta are strongly interacting 
with the CSM, at least during the first few months.

\begin{figure}
\includegraphics[width=8.6cm]{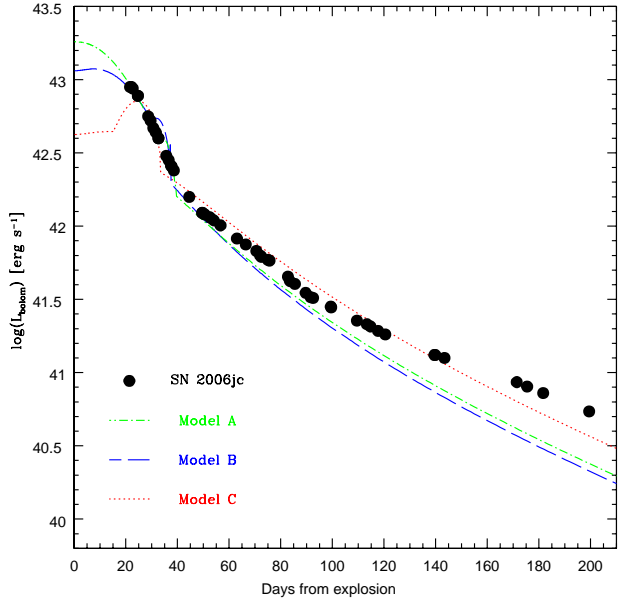}
\caption{Comparison between the bolometric light curve of SN~2006jc (filled black points) 
and the models A (large initial radius, small mass, low energy; green dot-dashed line), 
B (intermediate radius, ejected mass and explosion energy; blue dashed line) and C
(small radius, massive ejecta, high explosion energy; red dotted line) described in the text.  
 \label{fig10}}
\end{figure}

We consider the case of the well-observed SN 2006jc as representative
for the entire class. Here we make use of black body luminosities
from \citet{seppo07} obtained by fitting two components (hot
photosphere and warm dust) to the optical to NIR spectral energy
distribution (SED) of the SN to construct another bolometric light curve for
SN 2006jc. 
At early times a significant contribution to the total luminosity is
expected to come from wavelengths shorter than $\sim$3500\AA, while at later
epochs most of the contribution is in the IR domain. Since there were no NIR
data for 2006jc earlier than $\sim$55 days, the black body luminosities were
obtained only for epochs later than this \citep{seppo07}. However, at
+55 days the blackbody bolometric luminosity was dominated (more than 90\%
of the total luminosity) by the hot component and was about 2 times higher
(mainly due to the UV contribution) than the uvoir luminosity calculated in
Section \ref{bolo} of this paper. Missing any information on the blackbody
bolometric luminosity at earlier epochs, we scale up the luminosity earlier 
than +55 days (see Section \ref{bolo}) by a factor 2. For epochs later than +55
days we scaled the uvoir light curve up to match the appropriate blackbody
luminosities. This provides
a rough estimate (probably a lower limit) to the early time bolometric
luminosity of SN 2006jc. The resulting bolometric light curve is shown in
Fig. \ref{fig10}, where it is compared with the models described below.

In our attempt to reproduce its peculiar evolution, we model the bolometric light 
curve of SN~2006jc using three different explosion scenarios. 
In one of the models (model C) we also take into account a significant contribution to the energy output 
from the recombination of a residual He envelope: this may be partly responsible for the 
asymmetric peak and the steep post-maximum luminosity decay.  
The He recombination is an additional ingredient to a more traditional, simplified treatment 
of the bolometric light curves of SNe~ Ib/c \citep{arne82}, in which the SN evolution can be schematically divided into the 
photospheric (diffusive) phase and the nebular (radioactive) phase.
The ejecta are assumed to be in homologous expansion and spherically symmetric.
During the photospheric phase, we also assume that the energy output comes from the internal thermal energy of the ejecta, 
the recombination of the residual He envelope and the energy produced by the $^{56}$Ni $\rightarrow$ $^{56}$Co $\rightarrow$ $^{56}$Fe
decay chain. During the nebular phase, the ejecta are optically thin and the SN luminosity is merely supported
by the energy deposition of $\gamma$-rays from the $^{56}$Co decay. In the models, the effects of 
incomplete $\gamma$-ray trapping are taken into account.
More details on the bolometric light curve modelling adapted to H-deprived core-collapse SNe can be found in \citet{arne82}, while 
the treatment of He recombination is suitably adapted from the analysis presented in \citet{zam03} for type II 
SNe with massive H envelopes.

\begin{table*}
\footnotesize
\caption{Physical parameters of models A, B and C presented in Fig. \ref{fig10} and described in more detail in the text. \label{tab4}}
\begin{tabular}{cccccccccccc}
\hline\hline
Models & R$_0$ (10$^{12}$ cm)& E$_0$ (10$^{51}$ erg) & v$_0$ (km s$^{-1}$) & M$_{ej}$ (M$_\odot$) & M$_{Ni}$ (M$_\odot$) \\
\hline
A & 25$\pm$7  & 0.5$\pm$0.2 & 8000$\pm$1000  & 0.6$\pm$0.1 & 0.25$\pm$0.05 \\
B & 4$\pm$2   & 3.0$\pm$0.5 & 14000$\pm$1000 & 1.0$\pm$0.2 & 0.40$\pm$0.05 \\
C &1.5$\pm$0.5&  16$\pm$5   & 17000$\pm$1000 & 4.5$\pm$1.0 & 0.25$\pm$0.05 \\
\hline
\end{tabular}
\end{table*}

Satisfactory fits to the post-peak bolometric light curve of SN~2006jc are obtained assuming JD$_{exp}$ = 2454000
as an indicative explosion epoch. 
 After 4 months, the bolometric light curve of SN 2006jc flattens, possibly because of the
 increasing contribution from the
ejecta-CSM interaction or contamination from the galaxy background, and deviates significantly from the models (Fig. \ref{fig10}).
Model A (dot-dashed green line in  Fig. \ref{fig10}), which has the biggest initial radius ($\sim$2$\times$10$^{13}$ cm), the lowest expansion
velocity (v$_0$ = 8000 km s$^{-1}$) 
and a very low ejecta mass, fits quite well the observed SN data at early times. In this case a low explosion energy ($<$10$^{51}$ erg) and
0.2-0.3M$_\odot$ of radioactive $^{56}$Ni are required. Such amount of $^{56}$Ni, which is surprisingly large compared with 
the small total ejected mass, is necessary to reproduce the SN light curve.
However, if the late-time light curve is contaminated by ejecta-CSM interaction, the ejected $^{56}$Ni mass might be significantly lower.
Model B (blue dashed line) has a smaller initial radius (R$_0$ = 4$\times$10$^{12}$ cm), moderate ejected mass (about 1M$_\odot$), large $^{56}$Ni mass
($\sim$0.4M$_\odot$) and a rather standard explosion energy (3$\times$10$^{51}$ erg). Finally, model C (red dotted line) 
is the one with the smallest initial radius (around 10$^{12}$ cm) and
the highest ejected mass ($\sim$5M$_\odot$). It requires a very high explosion energy 
(about 10$^{52}$ erg) and 0.25M$_\odot$ of $^{56}$Ni. In this model, 
the recombination of the He envelope appears to  significantly affect the early-time light curve,
producing a peak at about 25 days after the explosion. 
Model A fits the early-time
bolometric curve of SN 2006jc  very well, 
although the initial radius is unreasonably large for a WR star, which is expected 
to be much more compact. Despite the three best fit models appear to be somehow degenerate 
in the post-peak phase, the late-time bolometric curve of SN 2006jc seems better matched by model C. 
The parameters adopted for the three models are listed in Table \ref{tab4}.

The set of models allows us to constrain the radioactive $^{56}$Ni mass to be in the range 0.2-0.4M$_\odot$.
However, a significant contribution to the luminosity from the ejecta-CSM interaction would imply a much
lower  $^{56}$Ni mass. 
The total ejected mass (and hence the progenitor mass) can not be well constrained, 
since the three models span a wide range of values between 0.5 and 5M$_\odot$.
The progenitor star, at the time of its explosion, could have been even more massive than $\sim$7M$_\odot$
\citep[this is the pre-SN mass roughly derived adding the 
mass of the remnant to the ejected mass of model C, and it is remarkably similar to that adopted by][]{tom07} 
because low-velocity material arising from a massive C-O mantle may well 
be present without significantly affecting the observed light curve.

Even though this preliminary data modelling cannot discriminate between high and low 
mass ejecta, the latter scenario seems inconsistent with other observational evidences. In particular,
the observed major eruption preceding SN~2006jc \citep{pasto07} and the large amount of material lost by the 
star before its core-collapse would suggest an originally very massive progenitor.
Future, detailed light curve and spectral modelling accounting also for the effects of the ejecta-CSM interaction, 
would provide more robust constraints on the physical parameters of SN~2006jc and similar events.

\section{Progenitors of Type Ibn Supernovae} \label{prog}

SN~2006jc has provided an excellent opportunity to study the nature of these objects, 
and to probe the nature of the progenitor stars. In \citet{pasto07} two different 
 scenarios were discussed for the precursor star of
SN~2006jc: a single WR star, originally very massive (up to 100M$_\odot$) and, 
alternatively, a double system formed 
by an LBV (responsible for the 2004 outburst) 
and a canonical WR star which exploded to produce SN~2006jc. 
In the following we will analyze in more detail these two 
scenarios, extending the discussion to the precursors of all SNe Ibc. 
 
\subsection{Single Massive Progenitor Scenario} \label{single}
A possible interpretation for the last period in the life of the progenitor of SN~2006jc 
is that an evolved WR star 
suffered a He-rich, luminous outburst in 2004 followed by the core-collapse 
two years later \citep{ita06,pasto07}. 
However this sequence of events is at odds with our current understanding of the 
late stages of evolution of the most massive stars. For example we have no evidence
locally that evolved WR stars produce sporadic He-rich mass ejections. There may be 
some evidence of a link between Ofpe/WN9 stars and LBVs but such objects still retain a 
He and H dominated envelope. The progenitor of SN~2006jc is not likely to have
been such a WN-LBV transition object when it underwent core-collapse, 
as there was no clear spectroscopic evidence of broad He I or H features 
which could be attributed to the high-velocity ejecta. 
The analysis of the spectral properties \citep{pasto07,fol07}
showed that the ejecta of SN~2006jc  were rich in intermediate-mass elements
like normal SNe Ic, and there was evidence of the presence of
a slowly moving, He-rich CSM.
Luminous outbursts, such as that observed in 2004 at M$_R \sim -14.1$ coincident with the position of SN 2006jc
(UGC 4904-V1),
are expected to be produced by proper LBV stars only \citep{dav97}
which are generally He and H rich, and indeed the outburst is accompanied by
the ejection of a large fraction of the outer envelope \citep{hum99}. 
However current stellar evolutionary theory would not predict that such $\eta$-Car type
objects would undergo core-collapse so close (2 years) to the LBV stage 
\citep{heg03,john06}. Instead, it would suggest a residual life time for the
star of another 10,000-100,000 years.

One would then be forced to conclude that 
the 2004 luminous outburst was produced by a WR star, which (almost) completely 
stripped its He envelope. The naked C-O WR star was still embedded in the  He-rich CSM 
at the time of the core-collapse. Since similar narrow He I features were observed in the spectra 
of all SNe Ibn, we can reasonably conclude that all their WR precursor stars likely suffered similar dramatic mass
loss episodes shortly before their death as SNe. While we cannot completely exclude the 
presence of broad He I components in the spectra of SN~2006jc, there is unequivocal evidence of such  
broad features only in the early-time spectra of SN~2000er and in the mid-age spectra of SN~2002ao. 
In these SNe, the narrow (1000-2000 km s$^{-1}$) circumstellar He I lines are found atop the broad ($\sim$5000 km s$^{-1}$)
He I components. This suggests that at 
least these two objects underwent core-collapse when the progenitor stars were in the transition between WN and WC 
phase \citep[for details on the physical sub-classification of WR stars, see e.g.][]{john06,crow07}.

Recently, \citet{smi06} and \citet{lang07} proposed that the bright type IIn SN~2006gy \citep[see also][]{ofek07} might be a pair-creation SN (PCSN), an exotic
event produced by the explosion of a very massive star \citep[more than 100M$_\odot$, ][]{heg02}. Though some of the expected 
properties of PCSNe are consistent also with the observables of type Ibn SNe (e.g. moderate to low
 metallicity 
of the host galaxies, slow observed spectral evolution, high peak luminosity, moderate expansion velocity of the ejecta), PCSNe are believed to produce
a much larger amount of $^{56}$Ni than observed in SN~2006jc and similar events. 

An alternative scenario, in good agreement with the observed properties of SN 2006gy and the brightest interacting SNe, 
was presented by \citet{woo07}. Some of these luminous events might originate from the collision between shells 
of material ejected by very massive stars (originally in the range 100-130M$_\odot$). The mechanism that 
leads to the ejection of many solar masses of the envelope, but not necessarily the explosion of the entire star, 
is the production of electron-positron pairs. After the first outburst, the stellar nucleus contracts, searching for 
a new phase of stable burning. A subsequent explosion may occur shortly after, ejecting several solar 
masses of material, which eventually collides with the material expelled in the previous outburst. According to  
\citet{woo07}, the luminosity of such events can be up to 10 times higher than that of a CC-SN. After the entire H
envelope is stripped away, also a sufficiently high mass of the remaining He core may generate multiple ejections of 
material \citep{woo07}. In light of all of this, the progenitor of SN 2006jc might have had a similar fate,
experiencing the last episode of such a sequence of luminous outbursts in 2004, followed by the 
core-collapse of the residual C-O core in 2006. 

\subsection{Binary (LBV+WR) Scenario} \label{double}
Although the single, massive progenitor scenario may offer some compelling explanations for most of the observed SN properties, 
it is at odds with current evolutionary theories in which no major LBV-like outburst is expected 
from a WR star. Even more surprising is that it occurred so close in time (only 2 years before) to the collapse of the stellar core.
There is also no observational evidence that any WR-type star has ever undergone such a luminous outburst.
All the bright outbursts from very massive stars have occurred in He+H-rich LBV stars \citep{hum99}.
These outbursts are likely accompanied by an ejection of $\sim$10 M$_\odot$ of material \citep{smi03}, which is uncomfortably
large to be consistent with an evolved WR progenitor.

An intriguing feature is the presence of a weak and narrow H$\alpha$ (v$_{FWHM} \sim$ 1000 km s$^{-1}$) in the late-time spectra of 
SN 2006jc \citep[see Fig. \ref{fig6} and][]{smi07}. This evidence, together with the detection of the luminous 
outburst in October 2004, led us to propose an alternative
scenario, i.e. a binary system composed of two massive stars at different evolutionary stages \citep{pasto07}. 
The pair could be formed by a typical LBV producing the 2004 transient and a more evolved WR star exploding as a supernova.
The LBV is losing its H envelope via recurrent outbursts like UGC 4904-V1, while the ejecta of SN 2006jc are colliding 
with the dense He-rich wind from the WR precursor star, possibly triggered by the binary interaction. 
Only once the SN ejecta reach the H-rich shell produced by the LBV, the
narrow H emission lines become visible in the SN spectra. 
While the overall spectral evolution is similar in the type Ibn SN group, there are significant differences in the strength of
the H$\alpha$ emission: it is clearly visible in the 1.5 month old spectrum of SN~2002ao shown by \citet[][see also our Fig. \ref{fig6b}]{fol07},  very weak in SN~2006jc
and undetected in SN~1999cq \citep{mat00} and in the very early spectra of SN~2000er. 
While the increasing strength of H$\alpha$ in the spectra of SN~2002ao would suggest a scenario analogous to that of SN~2006jc,
this is not necessarily true in the cases of SNe 1999cq and 2000er, which do not show any evidence
of narrow H lines in their spectra.

The probability of discovering binary systems composed by massive stars evolving with the same time scales might appear quite low.
However, massive binary systems in which one component is a WR 
star are frequently found in the Galaxy and the Local Group. In fact, a system similar to the putative precursor binary producing SN~2006jc
is well known in the Small Magellanic Cloud: HD5980 \citep[][and references therein]{koe04}.
The system consists of a primary massive star of about 50M$_\odot$ suffering episodic LBV outbursts suggesting
that the star is on its way to evolve into a He-rich WR star \citep{bar95,koe94,koe95}, 
and a more evolved eclipsing companion already beyond the LBV phase and identified as a $\sim$28M$_\odot$ WR star \citep{koe02}. 
The system might be even more complex, possibly hosting a third (O4-O6 type) 
star, gravitationally bound to the system but with a very long orbital period \citep{nie88,bre91,koe02}.
However, binary systems consisting of a WR star with a massive companion which may span a large range 
of evolutionary stages (main sequence O-type stars, red supergiants, LBVs or other WR stars) are quite common, e.g. 
systems with a WR plus a massive main sequence O-type star \citep{vdh01,sg01}, or WR binaries \citep{con89,foe03a,foe03b}.
Moreover, \citet{eld07} found that very massive stars which are thought to produce type
Ic SNe are more likely to have a binary companion, possibly an LBV, than lower mass stars.
In fact, assuming a flat secondary mass distribution,  \citet{eld07} estimated that  60 percent of the companions of
200M$_\odot$ primary SN precursors in binary systems  are post-main sequence objects, and therefore potentially LBVs. 
For a 100M$_\odot$ star the percentage decreases to 40 percent, while  for a 50M$_\odot$ star, it is only 20 percent.
With this evidence, a SN explosion of a WR star whose ejecta may at some stages collide with CSM released by 
the companion star might not be an unrealistic event.

\section{Frequency of Type Ibn Supernovae} \label{rate}

So far, only the 4 objects described in this paper \citep[plus maybe the transitional SN 2005la,][]{pasto07b} 
can be properly classified as type Ibn SNe. Since none were discovered before 1999, 
one may raise the question if these events are intrinsically rare or, perhaps, more 
common but misclassified in the past because of the inadequate spectral 
and photometric monitoring.
Although these SNe have unique characteristics, we can suggest only one object 
discovered before 1999 that may have some observed properties in common with SNe Ibn: the historical 
SN 1885A. 

\subsection{The Enigmatic SN 1885A (S Andromedae)} \label{SAn}

\begin{figure}
\includegraphics[width=8.9cm]{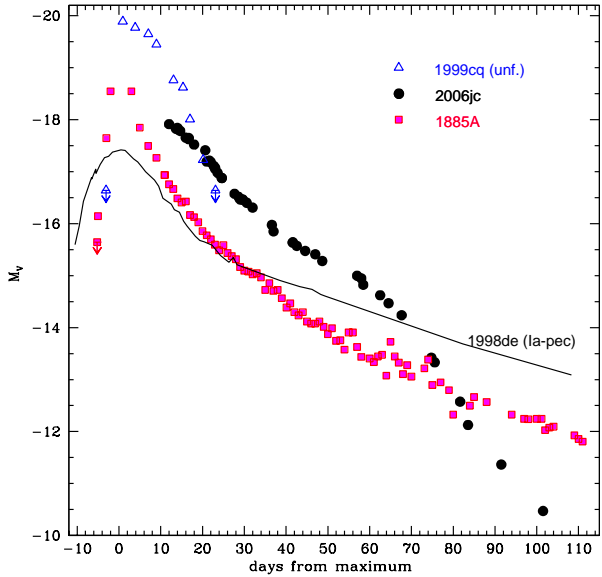}
\caption{V-band absolute light curve of SN 1885A \protect\citep{deva85} compared with those of SN 2006jc and the low-luminosity 
SN Ia 1998de \protect\citep{mod01,jha06}. The early, unfiltered  light curve of 1999cq \protect\citep{mat00} is also shown.\label{fig8}}
\end{figure}

The light curve of SN 1885A (also known as variable S Andromedae) was 
characterized by a  surprising, rapid rise to peak similar to that observed in SNe Ibn (see Fig. \ref{fig8}). 
This  enigmatic SN event 
 has been considered up to now as a peculiar 1991bg-like type Ia SN 
\citep[e.g.,][]{van94}. However, its light curve is different from that of any other SN Ia observed so far, having an 
asymmetric shape and an extremely fast post-max decline (Fig. \ref{fig8}). 
For the visual light curve of SN 1885A a $\Delta m_{15}(V)$  = 2.28 mag was estimated by \citet{van02}. 
This is much more than the V band post-maximum decline of the under-luminous type Ia
SN 1998de \citep[$\Delta m_{15}(V)$ = 1.31 mag,][]{mod01}. Using a relationship between colour and decline rate 
for underluminous 1991bg-like SNe Ia \citep[see also][]{gar04}, \citet{van02} obtained 
a value of $\Delta m_{15}(B)$ = 2.21 mag, which is not only inconsistent with that of a normal SN Ia, but
also much higher than any 1991bg-like event observed so far, being in 
the range between 1.6 and 2 \citep{gar04,tau07}. In addition, even with negligible extinction, SN 1885A is more 
luminous than the 1991bg-like SNe \citep[see e.g. SN 1998de in Fig.\ref{fig8},][]{mod01,jha06}. 
Some observed characteristics of the light curve of SN 1885A mimic those of SN~2006jc and similar objects, 
although its luminosity appears to be lower than those of the four events discussed here.  However, 
the host galaxy reddening is poorly known, and a significant internal extinction would make 
SN 1885A brighter and bluer. The pre-maximum rise of its light curve is extremely fast: the SN brightened by 3 magnitudes in 3-4 days \citep{van02}. Such a fast
luminosity evolution is comparable to that of the SNe of our sample, as tentatively determined from the epochs 
of pre-discovery limits and the earliest detections. 

So, can SN~1885A be considered as a nearby, 
historical analogue of SN~2006jc? There is some other evidence that might favour this scenario.
A historical spectrum of SN 1885A is described by \citet[][and references therein]{deva85}. 
The spectrum shows a continuum \citep[blue, according to][]{sher86} with possible absorption features at about 4500\AA~and 5700\AA, and 
several bright bumps that have been interpreted as emission features. In particular the strongest of them is
at about 5880\AA~ which was noted in all historical observations reported by  \citet{deva85}. This is
quite consistent with the He I 5876\AA~emission. 
Moreover,  \citet{bra87} classified SN~1885A as a peculiar type I SN (possibly of type Ib) remarking 
on the absence of prominent absorption features at  about 6150\AA. Hence the Si II 6355\AA ~absorption component 
typical of SNe Ia is probably missing. 

Despite this, a number of papers describing ground based and HST narrow-band imaging of the SN remnant  
\citep{fesen89,fesen99,ham00,fesen07} argue in favour of a type Ia 
nature for SN~1885A. All these studies show the 
remnant of SN 1885A to have spherical symmetry and to be detected in absorption against the luminous bulge of the host 
galaxy (M31), because of the prominent Ca I, Ca II, Fe I and Fe II absorption lines characterizing the remnant. 
Abundance arguments, the symmetric shape of the remnant and the lack of major star formation in the bulge
of M31 would all support a thermonuclear explosion scenario.
However the abundances predicted in these papers and the moderate ejected $^{56}$Ni mass \citep{fesen07}
are not inconsistent with those of luminous H-deprived core-collapse SNe. There is some evidence that
massive stars which explode as core-collapse SNe may sometimes be hosted in SO and early spiral galaxies, like the extremely luminous SN~2006gy \citep{smi06} 
and the ultra-faint SN IIP M85-OT2006 \citep{pasto07d}, although we would admit that 
 the true nature of these transients is currently being debated 
\citep{ofek07,kul07,rau07}. Despite the decades of studies of SN 1885A, the real nature of this event
still remains a puzzle.

\subsection{Type Ibn SN rate}

\begin{figure}
\includegraphics[angle=270,scale=0.35]{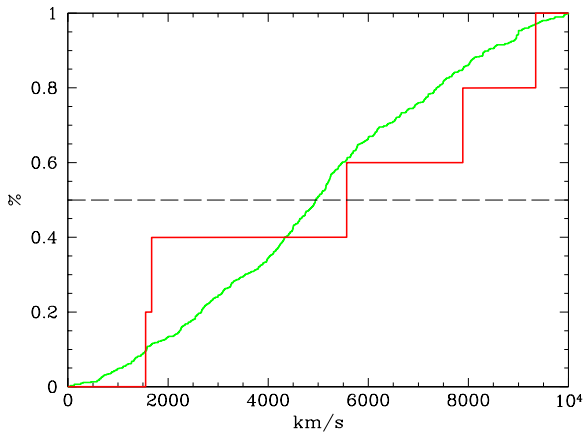}
\caption{Cumulative distribution of the radial velocities of the host galaxies of all CC SN
types (green line) and type Ibn SNe only (red line), within cz $\sim$ 10000 km s$^{-1}$,
as derived from the Asiago Supernova Catalogue. \label{fig11}}
\end{figure}

Some basic properties of SN progenitors (such as their interval of initial
mass) can be estimated through the measurement of the local SN rates
\citep[e.g.][]{cap97,cap99,mann05,mann06,bot07}. The measurement of
the SN rate is based on the ``control time'' methodology, developed by
Zwicky in the 1930s which requires intensive observational campaigns on
selected galaxies and the use of light curve templates for each SN
type. For the case of SNe Ibn the estimate of their frequency of
occurrence is complicated by the fact that the total amount of the
control time is not accessible and the available statistics are made
up of just a handful of objects. However, we can attempt to get around
these drawbacks by assuming that all galaxies within a given volume
have been monitored consistently and with similar control times. 
If this is assumed true, a rough estimate of the
rate of SNe Ibn can be derived after re-scaling the rate of core-collapse SNe (CC SNe)
(expressed in terms of SNuB or SNuM\footnotemark )
\footnotetext{SNuB is the number of SNe per century per 10$^{10}$L$_\odot$ in the B band, 
while SNuM is the number of SNe per century per 10$^{10}$M$_\odot$ of stellar mass.}
with the ratio
(SNe Ibn)/(CC SNe). The distance limit has been set after
comparing the cumulative frequencies of the distribution of core-collapse SNe and
SNe Ibn, which result to be statistically indistinguishable within cz $\sim$
10000 km s$^{-1}$ (see Fig. \ref{fig11}, Kolmogorov-Smirnov probability P=0.74). 
We also note that the first SN which can be properly classified
as a type Ibn event has been discovered in 1999. 
Though intrinsically rare, it is very likely that some
other SNe Ibn might have occurred before 1999, but probably
mis-classified, due to poor spectroscopic follow-up, as type IIn or Ib/c events
(or even as peculiar type Ia SNe, see Section \ref{SAn}). 
Therefore a temporal cut-off around 1999 has been applied to the
upgraded version of the Asiago SN catalogue\footnotemark ~\citep[see also][]{bar99}.
\footnotetext{http://web.oapd.inaf.it/supern/cat/}
Our analysis finds that, after 1999,
700 SNe have been classified as type II or Ib/c SNe within
cz = 10000 km s$^{-1}$. From this, and considering also the transitional SN~2005la \citep{pasto07b}
as a type Ibn SN event, we derive a ratio (SNe Ibn)/(CC SNe) $\sim$ 10$^{-2}$
which corresponds to 0.005-0.01 SNuB and 0.002-0.008 SNuM after
assuming an average Hubble type for the parent galaxies of SNe Ibn
ranging from Sa to Scd (see col. 5 in Table \ref{tab1}).

These values are remarkably consistent with the ratio (SNe Ibn)/(CC SNe)
derived taking into account all the nearby (v$_{rec} \leq$ 2000 km s$^{-1}$)
core-collapse SN events discovered during the period 1999-2006. With a total sample of 76 core-collapse SNe,
\citet{sma07} found (SNe Ibn)/(CC SNe) $\approx$ 2.5 $\times$ 10$^{-2}$.

\section{Summary} \label{end}

We have presented new data for three type Ibn SNe, i.e. SN~2000er, SN~2002ao and SN~2006jc. 
In particular, the early-time data of SN~2000er 
show, for the first time, the early spectroscopic behaviour of a type Ibn event. 
Early-time spectra of SN~2000er show relatively broad wings in the He I lines, with additional  
much narrower components. The detection of broad He I spectral features is unequivocal 
evidence for the presence of  He also in the SN ejecta, not only in the CSM.
Therefore, a residual envelope of He might be present in the progenitors of some SNe Ibn. 
Finally, the very late spectra of SN~2006jc still show strong narrow circumstellar He I lines,
together with an emerging H$\alpha$ likely produced in another CSM region. 

This data set allows us to highlight some key points concerning this new class of SNe, since  
we have now a better picture of the overall spectro-photometric evolution of SNe Ibn, 
which is essential to better constrain the characteristics of this class:
i) SNe Ibn show a surprisingly high degree of homogeneity, which is unexpected
 if the ejecta were strongly interacting with surrounding CSM;
ii) the modelling of the bolometric light curve of SN~2006jc (assuming no ejecta-CSM interaction) suggests
that an amount of 0.2-0.4M$_\odot$ of $^{56}$Ni was ejected. The mass of $^{56}$Ni could even be slightly 
higher for the brightest SNe Ibn (e.g. SN~1999cq).
However, we cannot exclude that the interaction between SN ejecta and CSM may power 
the light curves of SNe Ibn to some degree, hence the $^{56}$Ni mass might be somewhat lower;
iii) the observed data can be modelled  without invoking extraordinarily massive ejecta;
iv) their intrinsic rarity (being $\sim$ 1$\%$ of all core-collapse SNe) is probably due to the fact that
they arise from a rather exotic progenitor scenario, 
consistent with either a post-LBV/early-WR channel of a very massive (60-100M$_\odot$) star, or a binary (or even multiple) 
system formed by a normal LBV plus a core-collapsing WR companion. 

\section*{Acknowledgments}

We are grateful to R. J. Foley, M. Ganashelingham,  W. Li, A. V. Filippenko, H. Kawakita, K. Kinugasa and T. Matheson for
giving access to their data of SN~2002ao, SN~2006jc and SN~1999cq. 
AP thanks S. Blinnikov, M. T. Botticella, R. Kotak, P. Lundqvist, W. P. S. Meikle, and F. Patat for useful discussions.
SM acknowledges financial support from the Academy of Finland (project: 8120503).
SB, EC and MT are supported by the Italian Ministry of Education via the
PRIN 2006 n.2006022731$\_$002.
MDV thanks the Astrophysics Research Centre of the Queen's University of Belfast, 
where part of this work was done, for its hospitality.
The paper is partly based on observations collected at the European Southern Observatory,
in the course of the programs 66.B-0271, 66.A-0703, 66.A-0413, 66.D-0683, 78.D-0246.
The paper is also based on observations collected at the 
Copernico 1.82m Telescope of the INAF-Asiago Observatory (Mt. Ekar, Italy), the 1m Telescope of the 
Wise Observatory (Tel-Aviv University, Israel), the 2.3m Bok Telescope and the Kuiper 1.55m Telescope 
of the Steward Observatory (Arizona, US), the Liverpool Telescope, the William Herschel Telescope and 
the 0.80m Telescope of the Istituto de Astrofisica de Canarias in La Palma (Spain).
We thank the support astronomers of the Liverpool Telescope 
for performing the follow--up observations of SN~2006jc. 
This research has made use of the NASA/IPAC Extragalactic
Database (NED) which is operated by the Jet Propulsion Laboratory,
California Institute of Technology, under contract with the National
Aeronautics and Space Administration. We also made use of the HyperLeda database.\\


\end{document}